\documentclass[12pt]{article}
\usepackage{amsmath}
\usepackage{caption}
\usepackage{hyperref}
\usepackage{authblk}
\usepackage{epstopdf}
\usepackage{cite}
\usepackage{float}
\usepackage{mathrsfs}
\usepackage{subcaption}
\usepackage{amssymb}
\usepackage{mathtools}
\usepackage{graphicx}
\usepackage{xcolor}
\usepackage{filecontents}
\usepackage[T1]{fontenc}

\definecolor{tab:red}{RGB}{200,0,0}
\definecolor{tab:blue}{RGB}{10,0,200}
\definecolor{tab:green}{RGB}{10,200,10}
\definecolor{tab:black}{cmyk}{0,0,0,1}
\definecolor{pink}{rgb}{1.0,0.75,0.8}
\definecolor{purple(x11)}{rgb}{0.41,0.21,0.61}
\definecolor{orange(webcolor)}{rgb}{0.98,0.6,0.01}
\definecolor{gray}{rgb}{0.5,0.5,0.5}
\definecolor{brown(traditional)}{rgb}{0.59,0.29,0.0}

\begin{document}
\nocite{*}
\baselineskip=18 pt
\title{A new class of traversable exponential wormhole metrics}
\author[1]{\small  Partha Pratim Nath}
\author[2]{\small  Debojit Sarma}

\affil[a]{\footnotesize Department of physics, Cotton University}
\affil[1]{\textit{phy1891005\_partha@cottonuniversity.ac.in}}
\affil[2]{\textit{sarma.debojit@gmail.com}}
\maketitle

\begin{abstract}
   In this work we have formulated a new class of traversable exponential wormhole metrics. Here initially we have considered a exponential wormhole metric in which the temporal component is an exponential function of $r$ but the spatial components of the metrics are fixed as a particular function $e^{\frac{2m}{r}+2\alpha r}$. Following that, we have constructed a generalised exponential wormhole metric in which the spatial component is an exponential function of $r$ but the temporal component is fixed as a particular function given by $e^{-\frac{2m}{r}-2\alpha r}$. Finally we have considered exponential metric in which both the temporal and spatial components are generalised exponential function of $r$. We have also studied some of their properties including throat radius, stability, energy conditions, examined singularity, the metric in curvature coordinates, effective refractive index, innermost stable circular orbit(ISCO) and photon sphere, Regge-Wheeler potential and determined the curvature tensor. The radius of the throat is found to be consistent with the properties of wormholes and donot contain any types of singularities, which are given by $r=m$, $r = \frac{-1+\sqrt{1+4\alpha m}}{2\alpha}$, $r=\frac{-1+\sqrt{1+8\alpha m}}{4\alpha}$, $r=m+\frac{1}{\alpha}$, $r=m+\frac{2}{\alpha}$...etc. Most interestingly, we find that their throat radius is same for the same spatial component and the same range of values of $m$. In addition to these they also violate Null Energy Condition(NEC) near the throat. These newly constructed metrics form a new class of traversable wormhole.
\end{abstract}


\section{Introduction}
\par Wormholes are very fascinating topic for physicists from the early 90's. Theoretically significant progress has been made on this topic, yet no physical evidence has been found. This concept appears in various research papers and is a staple of science fiction. From the theoretical point of view, they are the solution to the Einstein's Field Equation. It is defined as a hypothetical tunnel connecting two different regions of spacetime. One important characteristics about wormholes is the wormhole throat, which can be defined as a two dimensional hypersurface of minimal area or the point where the radius is minimum \cite{visser1997geometric, hochberg1997geometric}. Traversability is also an important property of wormhole. If anything that enters through one side of the wormhole can exit through the other, the wormhole is traversable. One of the most important characteristics of a traversable wormhole is the absence of a horizon, because the presence of the horizon would prevent the two-way travel through the wormhole (Morris, Thorne \cite{morris1988wormholes}, presented nine desirable properties for a traversable wormhole, some are mandatory for the existence, while others are chosen to make the calculations simple). The existence of a traversable wormhole is possible if certain energy conditions such as the null energy condition(NEC) and the averaged null energy conditions(ANEC) are violated \cite{visser1997geometric, morris1988wormholes, morris1988wormholes1, hochberg1997self, friedman1993topological}. The coupling between spin and torsion provides more physically attainable sources for Morris-Thorne wormholes\cite{di2017spin}.
\par A considerable number of wormhole metrics have been constructed in the past. The very first wormhole solution was studied by Ludwig Flamm in Einstein gravity \cite{flamm1916beitrage,flamm2015republication}, but his solution was unstable. In 1928 Hermann Weyl \cite{scholz2001hermann} proposed a wormhole hypothesis of matter in connection with mass analysis of electromagnetic field energy. However, Weyl use the term 'one-dimensional tube' instead of the term 'wormhole'. Later Einstein and Rosen \cite{einstein1935particle} constructed a solution known as Einstein-Rosen bridge and they studied the solution in details. These wormholes were not stable \cite{fuller1962causality}. In 1988 Thorne and his graduate student Morris \cite{morris1988wormholes} constructed a wormhole with a throat and two mouths and introduced static traversable wormholes. Wheeler coined the term 'wormhole' and later his solutions were transformed into Euclidean wormholes by Hawking\cite{hawking1988wormholes} and others. The possible existence of wormholes draws attention to many questions like causality, energy conditions, geometric structure, topology of spacetime etc. 
\par In this paper we are interested to discuss some exponential metrics and will attempt at constructing  a new class  of traversable exponential wormhole metric. The notable exponential metric has the standard form 
\begin{equation}\label{(1)}
ds^2 = - \exp\left (-\frac{2m}{r} \right )dt^2 +\exp\left (\frac{2m}{r}\right )dr^2 + \exp\left (\frac{2m}{r}\right )[r^2 d\theta^2 + r^2 sin^2\theta d\phi^2].
\end{equation}
For more than 60 years \cite{yilmaz1958new, yilmaz1973new, clapp1973preliminary, fennelly1977nonviability, misner1999yilmaz, robertson1999x, ben2007relativistic, martinis2010exponential, ben2011some, svidzinsky2017vector, robertson2016meco, boonserm2018exponential}, since 1958, this exponential metric has been studied by many researchers. This metric has some attractive features that it passes almost all of the standard lowest order($\frac{2m}{r}<<1$) weak field test of General Relativity. But strong field behaviour($\frac{2m}{r}=1$)and medium field behaviour($\frac{2m}{1}>>1$) are very different \cite{boonserm2018exponential}. Since the temporal component $g_{tt}\neq0$, this exponential metric has no horizons, so it cannot be treated as a blackhole. This exponential metric has a wormhole throat at $r=m$ \cite{boonserm2018exponential} and describes a traversable wormhole in the sense of Morris and Thorne \cite{morris1988wormholes, morris1988wormholes1, visser1989traversable, visser1989traversable1, hochberg1997geometric, visser1995lorentzian, bhawal1992lorentzian, maldacena2017diving, lobo2005phantom, willenborg2018geodesic}.
\par Later Martinez and Nozawa \cite{martinez2021static} classified all the spherically symmetric solutions into three distinct classes namely Fisher class, Ellis-Gibbons class and Ellis-Bronnikov class. The general metric for Ellis-Gibbons class is given by
\begin{equation}\label{(EG)}
ds^2=-\exp\left(-\frac{M}{r^{d-3}}\right)dt^2+\exp\left(\frac{M}{(d-3)r^{d-3}}\right)[dr^2+r^2d\Sigma^2_{k=1,d-2}]
\end{equation}
This is a family of phantom-scalar solution and referred to as the "exponential metric". When $d=4$ and $d\Sigma^2_{k=1,2}$ being a metric of sphere, the above equation reduces to the metric Eq.(\ref{(1)}). They further demonstrated that for the metric Eq.(\ref{(EG)}) there exist spacetime points where some Riemann tensor components in a frame which is parallely propagated along the radial null geodesics diverges, that is they possess naked p.p. curvature singularities. According to them these solutions fail to describe regular wormhole solutions.
\par Here in this work we have developed a series of traversable exponential wormhole metrics in isotropic coordinates. Where the temporal and the spatial components are exponential functions of $r$. We have also studied some properties such as the throat radius, energy conditions, examined singularity, gravitational lensing properties such as effective refractive index, photon sphere, ISCO, Regge-Wheeler potential etc. of these newly constructed wormhole metrics. They donot possess any kind of singularities. Though we have named them as exponential wormhole metric, but they are different from the conventional exponential metric Eq.(\ref{(1)}) which falls into Ellis-Gibbons class. Interestingly we find that these newly constructed metrics form a differently new class of traversable wormhole metric. Our main motivation to formulate this new class of metrics is that we can build a theoretical model having some interesting phenomenological implications. One of the most attractive feature of these metrics is that they are traversable in nature, where interestingly on the otherside of the wormhole throat time is observed to be slowed down. It also implies that unstable photon orbits and ISCO still exist with slight shift from the position in Schwarzschild space-time and Regge-Wheeler potentials can also be studied.
\par The paper is assembled as follows. In Sec.II, we have constructed a series of traversable exponential wormhole metrics in isotropic coordinates and also checked their throat radius, Einstein tensors and Null Energy Condition. Stability conditions are studied in Sec.III. In Sec.IV, we have transformed these newly constructed metrics to curvature coordinates and calculated the throat radius in terms of curvature coordinates, followed  by the evaluation of the non zero curvature tensors and examine singularity in Sec.V. In Sec.VI, we have studied the lensing properties of these metrics by evaluating the effective refractive indices. In Sec.VII, we have calculated the radius of Photon Sphere and Innermost Stable Circular Orbit(ISCO) followed by the Regge-Wheeler potentials in Sec.VIII. Finally we present results and discussion in the Sec.IX.
 
\section{Construction of traversable exponential wormhole metric}

Here, we have a space time metric in exponential form by slightly modifying the conventional exponential wormhole metric Eq.(\ref{(1)})
\begin{equation}\label{(2)}
ds^2 =  - \exp\left (-\frac{2m}{r} \right )dt^2+\exp\left (\frac{2m}{r} + 2\alpha r\right )dr^2+
\exp \left (\frac{2m}{r} + 2\alpha r \right )[r^2 d\theta^2 + r^2 sin^2\theta d\phi^2],
\end{equation}
where $\alpha$ is a constant which has a dimension inverse of length. $\alpha$ effects the radius of curvature of the wormhole. All the metrics that we consider in this work follows the standard form of Isotropic metric, which is given by,
\begin{equation}\label{isotropic}
ds^2 = -a(r)^2dt^2 + b(r)^2[dr^2 + r^2(d\theta^2 + sin^2\theta d\phi^2)],
\end{equation}
where $a$ and $b$ are functions of the radial coordinate.
\\
The exponential metric[{\ref{(20)}}] has no horizons as $g_{tt}\neq0$, so it is not a Blackhole. Now,we consider the area of the spherical surfaces of constant $r$ coordinate:
\begin{equation}\label{(3)}
A(r)=4\pi r^2 \exp\left(\frac{2m}{r} + 2\alpha r\right).
\end{equation}
Therefore
\begin{equation}\label{(4)}
\frac{dA(r)}{dr}= 8\pi r \exp\left (\frac{2m}{r} + 2\alpha r\right )-8\pi m \exp \left (\frac{2m}{r} + 2\alpha r\right )
+8\alpha \pi r^2 \exp \left(\frac{2m}{r} + 2\alpha r\right ).
\end{equation}
Equating Eq(\ref{(4)}) with $0$, 
we obtain
\begin{equation}\label{(5)}
r=r_0 = \frac{-1+\sqrt{1+4\alpha m}}{2\alpha}.
\end{equation}
if $0<m<<\frac{1}{4\alpha}$, then
\begin{equation}\label{(6)}
r=r_0=m
\end{equation}
For this value of $r$, we have $\frac{d^2 A(r)}{dr^2}=8\pi e^{(2\alpha m+2)}[2\alpha^2 m^2+1] >0$
\\i.e., the area is a concave function of the $r$ coordinate, and has a minimum at $r=r_0=m $.
So the term $r_0=m$ can be considered as the radius of the throat, where it satisfies the "flare out" condition and all metric components are finite at $r=m$ and the diagonal components are non-zero. From this we can certainly say that the surface $r=m$ for $0<m<<\frac{1}{4\alpha}$ is a wormhole throat for our metric.
\\From the explicit forms of the non-zero Einstein tensor components(in the units of $c=1,G=1$), we get,

\begin{equation}\label{(10)}
p_r (r) =\exp\left(-\frac{2m}{r} -2\alpha r\right)\frac{(-m^2+\alpha r^3(2+\alpha r))}{8\pi r^4},
\end{equation}
\begin{equation}
\rho(r)=-\exp\left(-\frac{2m}{r} -2\alpha r\right)\frac{(m^2-2mr^2\alpha+r^3\alpha(4+\alpha r))}{r^4}
\end{equation}
and
\begin{equation}
p_t(r)=\exp\left(-\frac{2m}{r} -2\alpha r\right)\frac{(m^2+\alpha r^3)}{r^4}.
\end{equation}
Where $\rho$, $p_r$ and $p_t$ stand for energy density, radial and tangential pressure respectively.\\
Therefore,
\begin{equation}\label{(12)}
p_r (r) + \rho (r) = \exp\left(-\frac{2m}{r} -2\alpha r\right)\frac{(-2m^2-2\alpha r^3+2\alpha m r^2)}{8\pi r^4}.
\end{equation}

Now if $0<m<<\frac{1}{4\alpha}$ at the throat, then
\begin{equation}\label{(13)}
p_r(r_0) + \rho (r_0) = -\frac{e^{(-2\alpha m-2)}}{4\pi m^2} < 0.
\end{equation}
The metric violates the NEC at the throat of the wormhole.
So, we can say that the throat is filled with the exotic matter for $0<m<<\frac{1}{4\alpha}$,
and the radius at the throat becomes $r_0 = m$.
\\
\subsection{Temporal component}  Now, let us consider a general exponential metric, in which the temporal part is a general function of $r$ in the form $e^{-\frac{2m}{r}-2\alpha^n r^n}$ and the spatial component is considered as $e^{\frac{2m}{r}+2\alpha r}$
\begin{equation}\label{(14)}
ds^2 = - \exp\left(-\frac{2m}{r}-2\alpha^n r^n\right)dt^2 + \exp\left(\frac{2m}{r} + 2\alpha r\right)dr^2 + \exp\left(\frac{2m}{r} + 2\alpha r\right)[r^2 d\theta^2 + r^2 sin^2\theta d\phi^2],
\end{equation}
where $n$ is an integer. 
Like earlier metric Eq.(\ref{(2)}), this metric also does not contain any horizon. Here the value of throat radius is $r=r_0 = \frac{-1+\sqrt{1+4\alpha m}}{2\alpha}$. If $0<m<<\frac{1}{4\alpha}$, then $r=r_0=m$.
\\Now, the explicit form of the nonzero Einstein tensor components are,
\begin{equation}\label{(15)}
G^r_r = \exp\left(-\frac{2m}{r} -2\alpha r\right)\frac{(-m^2+\alpha r^3(2+\alpha r)-2nr^{n+1}\alpha^n(-m+r+\alpha r^2))}{r^4},
\end{equation}
\begin{equation}\label{(16)}
G^t_t = \exp\left(-\frac{2m}{r} -2\alpha r\right)\frac{(m^2-2mr^2\alpha+\alpha r^3(4+\alpha r))}{r^4}
\end{equation}
and
\begin{equation}\label{(17)}
G^{\theta}_{\theta} = G^{\phi}_{\phi} = \exp\left(-\frac{2m}{r} -2\alpha r\right)\frac{(m^2+\alpha r^3-nr^{n+1}(2m+nr)\alpha^n+n^2\alpha^{2n}r^{2n+2})}{r^4}.
\end{equation}
In the units of $c=1,G=1$, we obtain
\begin{equation}\label{(18)}
p_r (r)=\exp\left(-\frac{2m}{r} -2\alpha r\right)\frac{(-m^2+\alpha r^3(2+\alpha r)-2nr^{n+1}\alpha^n(-m+r+\alpha r^2))}{8\pi r^4}
\end{equation}
and
\begin{equation}\label{(19)}
\rho (r) =  -\exp\left(-\frac{2m}{r} -2\alpha r\right)\frac{(m^2-2mr^2\alpha+\alpha r^3(4+\alpha r))}{8\pi r^4}.
\end{equation}
Now,
\begin{equation}\label{(20)}
p_r (r) + \rho (r) =\exp\left(-\frac{2m}{r} -2\alpha r\right)\frac{(-2m^2+2(m-r)\alpha r^2-2nr^{n+1}\alpha^n(-m+r+\alpha r^2))}{8\pi r^4}.
\end{equation}
At the throat region, if  $0<m<<\frac{1}{4\alpha}$, then
\begin{equation}\label{(21)}
p_r(r_0) + \rho (r_0) = -\frac{e^{(-2\alpha m-2)}(1+n\alpha^{n+1}m^n)}{4\pi m^2} < 0,
\end{equation}
i.e., it violates the NEC. So we can conclude that the throat is filled with some exotic matter which violates the NEC.
\\
\\Now, combining the metrics Eq.(\ref{(2)}) and Eq.(\ref{(14)}), we get a exponential wormhole metric. While on changing the values of $n$, we can get different wormhole metrics. Again if $0<m<<\frac{1}{4\alpha}$, the throat radius is $r=r_0 = m$. Interestingly we observe that the throat radius remains same for fixed values of $m$ and $\alpha$ irrespective of the values of $n$. 
\\
\\Again, let us consider an another exponential metric in which the power of the exponential term of the temporal component is the sum of two $"r"$ with different powers,
\begin{equation}\label{(22)}
ds^2 = -\exp\left(-\frac{2m}{r}-2(\alpha^n r^n+\alpha^l r^l)\right)dt^2 + \exp\left(\frac{2m}{r}+2\alpha r\right)[dr^2 + r^2 d\theta^2 + r^2 sin^2\theta d\phi^2]
\end{equation}
\begin{equation}\label{(23)}
\implies ds^2 = -\exp\left(-\frac{2m}{r}-2\alpha^n r^n-2\alpha^l r^l\right)dt^2 + \exp\left(\frac{2m}{r}+2\alpha r\right)[dr^2 + r^2 d\theta^2 + r^2 sin^2\theta d\phi^2]
\end{equation}
where $n$ and $l$ are two different integers.
\\Again, this metric does not have any horizon as $g_{tt}\neq0$ . For this metric the value of throat radius is again $r=r_0 = \frac{-1+\sqrt{1+4\alpha m}}{2\alpha}$ and if $0<m<<\frac{1}{4\alpha}$, then $r=r_0=m$. The values of $n$ and $l$ have no impact on the throat radius.
\\The explicit forms of the nonzero Einstein tensor components are
\begin{align}\label{(24)}
\begin{split} 
G^r_r={}&\frac{\exp\left(-\frac{2m}{r}-2\alpha r\right)}{r^4}[-m^2+\alpha r^3(2+\alpha r)+2mr(l\alpha^l r^l+n\alpha^n r^n)-2r^2(1+\alpha r)(l\alpha^l r^l\\
&+n\alpha^n r^n)],
\end{split}
\end{align}
\begin{equation}\label{(25)}
G^t_t= \exp\left(-\frac{2m}{r}-2\alpha r\right)\frac{[m^2-2m\alpha r^2+\alpha r^3(4+\alpha r)]}{r^4}
\end{equation}
and
\begin{align}\label{(26)}
\begin{split} 
G^{\theta}_{\theta} = G^{\phi}_{\phi}={}&\frac{\exp\left(-\frac{2m}{r}-2\alpha r\right)}{r^4}[m^2+ \alpha r^3-2mr(l r^l \alpha^l +n r^n \alpha^n)+r^2(2lnr^{n+l} \alpha^{n+l}+l^2 r^l \alpha^l\\
&(-1+r^l \alpha^l)+n^2 r^n \alpha^n (-1+r^n \alpha^n))],
\end{split}
\end{align}
which leads to (in the units of $c=1$ and $G=1$),
\begin{align}\label{(27)}
\begin{split}
p_r (r) ={}&\frac{\exp\left(-\frac{2m}{r}-2\alpha r\right)}{8\pi r^4}[-m^2+\alpha r^3(2+\alpha r)+2mr(l\alpha^l r^l+n\alpha^n r^n)-2r^2(1+\alpha r)(l\alpha^l r^l\\
&+n\alpha^n r^n)],
\end{split}
\end{align}
\begin{equation}\label{(28)}
\rho (r) = - \exp\left(-\frac{2m}{r}-2\alpha r\right)\frac{[m^2-2m\alpha r^2+\alpha r^3(4+\alpha r)]}{8\pi r^4}
\end{equation}
Now we have,
\begin{align}\label{(29)}
\begin{split}
p_r (r) + \rho (r) ={}&\frac{\exp\left(-\frac{2m}{r}-2\alpha r\right)}{8\pi r^4}[-2m^2+2(m-r)\alpha r^2-2lr^{l+1} \alpha^l(-m+r+\alpha r^2)-\\
&2nr^{n+1} \alpha^n(-m+r+\alpha r^2)].
\end{split}
\end{align}
If $0<m<<\frac{1}{4\alpha}$, we get the throat radius $r=r_0=m$ and from Eq.(\ref{(29)}) we obtain
\begin{equation}\label{(30)}
p_r(r_0) + \rho (r_0) = - \frac{\exp\left(-2-2\alpha m\right)}{4\pi m^2}[1+l \alpha^{l+1} m^{l+1}+n \alpha^{n+1} m^{n+1}]<0,
\end{equation}
which shows that the metric violates the NEC at the throat. 
\\
\\In this way, we can make a traversable exponential wormhole metric of the following form,
\begin{align}\label{(31)}
\begin{split}
ds^2 ={}& -\exp\left(-\frac{2m}{r}-\alpha^n r^n-\alpha^l r^l-\alpha^k r^k-.......\alpha^q r^q\right)dt^2 + \exp\left(\frac{2m}{r}+2\alpha r\right)[dr^2 +\\
& r^2 d\theta^2 + r^2 sin^2\theta d\phi^2]
\end{split}
\end{align}
or
\begin{equation}\label{(32)}
ds^2 = -\exp\left(-\frac{2m}{r}-\sum_{\beta=0}^{\beta=n}\alpha^{\beta}r^{\beta}\\\right)dt^2 + \exp\left(\frac{2m}{r}+2\alpha r\right)[dr^2 + r^2 d\theta^2 + r^2 sin^2\theta d\phi^2],
\end{equation}
where the powers $n$, $l$, $k$....$q$ etc are integers. We can also show that the metrics Eq.(\ref{(31)}) and Eq.(\ref{(32)}) also contain a wormhole throat at $r=r_0=m$, for $0<m<<\frac{1}{4\alpha}$. So all the exponential metrics that can be formed from Eq.(\ref{(31)}) and Eq.(\ref{(32)}) would have the same throat radius $r=\frac{-1+\sqrt{1+4\alpha m}}{2\alpha}$ and if $0<m<<\frac{1}{4\alpha}$, then the throat radius will come out as $r=r_0=m$. The values of $n$, $l$, $k$ etc donot have any impact on the throat radius in the case of temporal metrics. Again all these metrics would violate the necessary Null Energy Condition (NEC) near the throat region.
\\
\\
\subsection{Spatial component} Now, let us consider a space-time metric in which we will keep the temporal component fixed as $\exp\left(-\frac{2m}{r}-2\alpha r\right)$, but the spatial component as a generalised form in terms of exponential function of $r$.
\begin{equation}\label{(37)}
ds^2 = -\exp\left(-\frac{2m}{r}-2\alpha r\right)dt^2 + \exp\left(\frac{2m}{r}+2\alpha^n r^n\right)[dr^2 + r^2 d\theta^2 + r^2 sin^2\theta d\phi^2],
\end{equation}
where $n$ is an integer. As $g_{tt}\neq 0$, so this metric also doesn't contain any singularity.
\\Now, from non-zero Einstein tensor and in the units of $c=1$ and $G=1$,
\begin{equation}\label{(38)}
p_r(r)=\frac{\exp\left(-\frac{2m}{r}-2\alpha^n r^n\right)}{8\pi r^4}[-m^2+2mr^2 \alpha+r^2(-2r\alpha +n^2 r^{2n} \alpha^{2n}-2nr^n \alpha^n(-1+\alpha r))]
\end{equation}
and
\begin{equation}\label{(39)}
\rho(r)=-\frac{\exp\left(-\frac{2m}{r}-2\alpha^n r^n\right)}{8\pi r^4}[m^2-2mnr^{n+1}\alpha^n +nr^{n+2}\alpha^n(2+n(2+r^n\alpha^n))].
\end{equation}
Now,
\begin{equation}\label{(40)}
p_r(r)+\rho(r)=\frac{\exp\left(-\frac{2m}{r}-2\alpha^n r^n\right)}{4\pi r^4}[-m^2+(m-r)\alpha r^2-nr^{n+1} \alpha^n(-m+r(n+\alpha r))].
\end{equation}
For this metric, the area of the spherical surface of constant $r$ coordinate is
\begin{equation}\label{(41)}
A(r)=4\pi r^2\exp\left(\frac{2m}{r}+2\alpha^n r^n\right).
\end{equation}
Then
\begin{equation}\label{(42)}
\frac{d(A(r))}{dr}=8\pi \exp\left(\frac{2m}{r}+2\alpha^n r^n\right)[r-m+n\alpha^n r^{n+1}].
\end{equation}
Now, equating it with zero, we get
\begin{equation}\label{(43)}
n\alpha^n r^{n+1}+r-m=0.
\end{equation}
Now, by solving the above equation we can get the value of radius at the throat. Here we will discuss some simple cases for some small values of $n$.
\\
\textbf{Case 1} Let us consider $n=0$, the metric Eq.(\ref{(37)}) becomes,
\begin{equation}\label{(44)}
ds^2 = -\exp\left(-\frac{2m}{r}-2\alpha r\right)dt^2 + \exp\left(\frac{2m}{r}+2\right)[dr^2 + r^2 d\theta^2 + r^2 sin^2\theta d\phi^2].
\end{equation}
For this metric, from Eq.(\ref{(43)}) , the radius at the throat will come out as $r=r_0=m$ (as at $r=m$, the $\frac{d^2(A(r))}{dr^2}>0$, that is the area has a minimum at $r=m$).  This $m$ can have any positive value.
\\Similarly, if we check the Null Energy Condition, then we will get from Eq(\ref{(40)}).
\begin{equation}\label{(45)}
p_r(r)+\rho(r)=\exp\left(-\frac{2m}{r}-2\right)\frac{[-m^2+(m-r)\alpha r^2]}{4\pi r^4}.
\end{equation}
And at the throat,
\begin{equation}\label{(46)}
p_r(r_0)+\rho(r_0)=-\frac{e^{-4}}{4\pi m^2}<0.
\end{equation}
Therefore, the NEC is violated for any value of $m$.\\ 
\textbf{Case 2} if $n=1$, then the metric Eq.(\ref{(37)}) takes the form,
\begin{equation}\label{(47)}
ds^2 = -\exp\left(-\frac{2m}{r}-2\alpha r\right)dt^2 + \exp\left(\frac{2m}{r}+2\alpha r\right)[dr^2 + r^2 d\theta^2 + r^2 sin^2\theta d\phi^2].
\end{equation}
This metric is also a special case of the general metric represented by Eq.(\ref{(14)}). For this metric, by solving Eq.(\ref{(43)}), we will get the radius at throat as $r=r_0 = \frac{-1+\sqrt{1+4\alpha m}}{2\alpha}$ and for the range $0<m<<\frac{1}{4\alpha}$, the radius at the throat becomes $r=r_0=m$.
\\Again from Eq.(\ref{(40)}), we will get, 
\begin{equation}\label{(48)}
p_r(r)+\rho(r)=\frac{\exp\left(-\frac{2m}{r}-2\alpha r\right)}{4\pi r^4}[-m^2+r^2(m-r)\alpha-\alpha r^2(-m+r(1+\alpha r))].
\end{equation}
At the throat, where $r=r_0=m$
\begin{equation}\label{(49)}
p_r(r_0)+\rho(r_0)=-\frac{e^{(-2\alpha m-2)}}{4\pi m^2}[\alpha^2 m^2+1]<0.
\end{equation}
Like earlier, we can again say that the Null Energy Condition (NEC) is violated in the throat region or the throat is filled with some exotic matter.\\
\textbf{Case 3} In this last case, we will consider the value of $n=-1$, then the metric of Eq.(\ref{(37)}) will take the form
\begin{equation}\label{(50)}
ds^2 = -\exp\left(-\frac{2m}{r}-2\alpha r\right)dt^2 + \exp\left(\frac{2m}{r}+\frac{2}{\alpha r}\right)[dr^2 + r^2 d\theta^2 + r^2 sin^2\theta d\phi^2].
\end{equation}
From Eq.(\ref{(43)}), the radius of the throat will come out as
\begin{equation}\label{(51)}
r=r_0= m+\frac{1}{\alpha}
\end{equation}
If $\alpha>> m$, then the radius becomes $r=r_0=m$. Again we can show that $p_r(r_0)+\rho(r_0)<0$.
\\In this way we can continue the process and can find different forms of exponential wormhole metric using different integral values of $n$.
\\
\\Again, we will consider an another exponential wormhole metric in which the spatial part is again a generalised function.
\begin{equation}\label{(52)}
ds^2 = -\exp\left(-\frac{2m}{r}-2\alpha r\right)dt^2 + \exp\left(\frac{2m}{r}+2\alpha^n r^n+2\alpha^k r^k\right)[dr^2 + r^2 d\theta^2 + r^2 sin^2\theta d\phi^2],
\end{equation}
where $n$ and $k$ are two different integers(they may be equal also). For this metric again $g_{tt}\neq0$, so it does not contain any type of singularity. Now from the non zero Einstein tensors and in the units of $c=1$ and $G=1$, we can write,
\begin{align}\label{(53)}
\begin{split}
p_r(r)={}&\frac{\exp\left(-\frac{2m}{r}-2r^k\alpha^k-2r^n\alpha^n\right)}{8\pi r^4}[-m^2+2mr^2 \alpha+r^2(-2r \alpha+2kr^k \alpha^k +k^2 r^{2k} \alpha^{2k}-\\
&2kr^{k+1} \alpha^{k+1}+n^2 r^{2n} \alpha^{2n}+2nr^n \alpha^n(1-\alpha r+kr^k \alpha^k))],
\end{split}
\end{align}
\begin{align}\label{(54)}
\begin{split}
\rho(r)={}&-\frac{\exp\left(-\frac{2m}{r}-2r^k\alpha^k-2r^n\alpha^n\right)}{8\pi r^4}[m^2-2mr(kr^k\alpha^k +nr^n\alpha^n)+r^2(n^2r^{2n}\alpha^{2n}\\& +2nr^n\alpha^n (1+n+kr^k\alpha^k)+kr^k\alpha^k(2+k(2+r^k\alpha^k)))]
\end{split}
\end{align}
and
\begin{align}\label{(55)}
\begin{split}
p_r(r)+\rho(r)={}&\frac{\exp\left(-\frac{2m}{r}-2r^k\alpha^k-2r^n\alpha^n\right)}{4\pi r^4}[-m^2+(m-r)\alpha r^2-kr^{k+1} \alpha^k(-m+r(k+\alpha r))\\
&-nr^{n+1} \alpha^n(-m+r(n+\alpha r))].
\end{split}
\end{align}
Now, the area of the spherical surface is
\begin{equation}\label{(56)}
A(r)=4\pi r^2\exp\left(\frac{2m}{r}+2\alpha^n r^n+2\alpha^k r^k\right),
\end{equation}
then, \\$\frac{d(A(r))}{dr}=8\pi\exp\left(\frac{2m}{r}+2\alpha^n r^n+2\alpha^k r^k\right)[r-m+n\alpha^n r^{n+1}+k\alpha^k r^{k+1}]$
\\equating it with zero, we will get,
\begin{equation}\label{(57)}
r-m+n\alpha^n r^{n+1}+k\alpha^k r^{k+1}=0.
\end{equation}
Now, by solving the above equation, we can find the radius at the throat. Here we will consider some cases for some values of $n$ and $k$.\\
\textbf{Case 1} If $n=0$ and $k=0$, then the metric Eq.(\ref{(52)}) takes the form,
\begin{equation}\label{(58)}
ds^2 = -\exp\left(-\frac{2m}{r}-2\alpha r\right)dt^2 + \exp\left(\frac{2m}{r}+4\right)[dr^2 + r^2 d\theta^2 + r^2 sin^2\theta d\phi^2].
\end{equation}
From Eq.(\ref{(57)}) we will get the value of throat radius as
\begin{equation}\label{(59)}
r=r_0=m,
\end{equation}
that is the radius at the throat is equal to $m$, for any positive values of $m$.
\\Again, the Null Energy Condition at the throat, from the Eq.(\ref{(55)}),
\begin{equation}\label{(60)}
p_r(r_0)+\rho(r_0)=-\frac{\exp\left(-6\right)}{4\pi m^2}<0.
\end{equation}
It shows that the metric violates NEC for $n=0$ and $k=0$.\\
\textbf{Case 2} If $n=k=-1$, the metric Eq.(\ref{(52)}) takes the form,
\begin{equation}\label{(61)}
ds^2 = -\exp\left(-\frac{2m}{r}-2\alpha r\right)dt^2 + \exp\left(\frac{2m}{r}+\frac{4}{\alpha r}\right)[dr^2 + r^2 d\theta^2 + r^2 sin^2\theta d\phi^2],
\end{equation}
the radius at the throat becomes,
\begin{equation}\label{(62)}
r=r_0=m+\frac{2}{\alpha}.
\end{equation}
If $\alpha>>m$, the throat radius becomes,
\begin{equation}\label{(63)}
r=r_0=m.
\end{equation}

Again, from Eq.(\ref{(55)}) , at the throat, we will get
\begin{equation}\label{(64)}
p_r(r_0)+\rho(r_0)<0,
\end{equation}
i.e., the NEC is again violated near the throat region.\\
\textbf{Case 3} If both $n=k=1$, the metric Eq.(\ref{(52)}) takes the form,
\begin{equation}\label{(65)}
ds^2 = -\exp\left(-\frac{2m}{r}-2\alpha r\right)dt^2 + \exp\left(\frac{2m}{r}+4\alpha r\right)[dr^2 + r^2 d\theta^2 + r^2 sin^2\theta d\phi^2].
\end{equation}
Now, this time by solving Eq.(\ref{(57)}), the radius at the throat comes out as,
\begin{equation}\label{(66)}
r=r_0=\frac{-1+\sqrt{1+8\alpha m}}{4\alpha}.
\end{equation}
If we take the range of $m$ as $0<m<<\frac{1}{8\alpha}$, then the throat radius becomes,
\begin{equation}\label{(67)}
r=r_0=m.
\end{equation}
Lastly, from the Eq.(\ref{(55)}), the NEC becomes,
\begin{equation}\label{(68)}
p_r(r_0)+\rho(r_0)=-\frac{e^{(-4\alpha m-2)}}{4\pi m^2}[1+2\alpha^2 m^2]<0.
\end{equation}
So, we have seen that the NEC is again violated near the throat region.
\\So in this way we can check for different values of $n$ and $k$, the metrics obtained are some exponential wormhole metrics. These metrics contain a stable throat as well as they violate the NEC near the throat region implying the presence of exotic or negative mass energy. 
\\Thus in this way, we can construct a traversable exponential wormhole metric as,
\begin{equation}\label{(69)}
ds^2 = -\exp\left(-\frac{2m}{r}-2\alpha r\right)dt^2 + \exp\left(\frac{2m}{r}+2\alpha^n r^n+2\alpha^k r^k+2\alpha^l r^l+...\right)[dr^2 + r^2 d\theta^2 + r^2 sin^2\theta d\phi^2],
\end{equation}
where $n$, $k$, $l$....etc are different integers. Here we will see that the radius of the throat will be different for different values of $n$, $k$, $l$...etc. 
\\
\subsection{Combination of temporal and spatial}
Here, in this part we will consider the both temporal and spatial components as a generalised exponential function of $r$, let us first consider a metric of the form as follows,
\begin{equation}\label{(70)}
ds^2 = -\exp\left(-\frac{2m}{r}-2\alpha^k r^k\right)dt^2 + \exp\left(\frac{2m}{r}+2\alpha^n r^n\right)[dr^2 + r^2 d\theta^2 + r^2 sin^2\theta d\phi^2].
\end{equation} 
As earlier, $n$ and $k$ are integers.
\\From Einstein non zero tensor and using the units of $c=1$ and $G=1$, we get
\begin{equation}\label{(71)}
p_r(r)=\frac{\exp\left(-\frac{2m}{r}-2\alpha^n r^n\right)}{4\pi r^4}[-m^2+2kmr^{k+1}\alpha^k +r^2(-2kr^k\alpha^k (1+nr^n\alpha^n)+nr^n\alpha^n (2+nr^n\alpha^n))],
\end{equation}
\begin{equation}\label{(72)}
\rho(r)=-\frac{\exp\left(-\frac{2m}{r}-2\alpha^n r^n\right)}{4\pi r^4}[m^2-2mnr^{n+1}\alpha^n +nr^{n+2}\alpha^n(2+n(2+r^n\alpha^n))]
\end{equation}
And
\begin{equation}\label{(73)}
p_r(r)+\rho(r)=\frac{\exp\left(-\frac{2m}{r}-2\alpha^n r^n\right)}{2\pi r^4}[-m^2+k(m-r)r^{k+1}\alpha^k-nr^{n+1}\alpha^n(-m+r(n+kr^k\alpha^k))].
\end{equation}
The area of the spherical surface is same as that of Eq.(\ref{(41)}). Now let us consider some values of $n$ and $k$\\
\textbf{Case 1} If both $n=k=0$, then metric takes the form,
\begin{equation}\label{(74)}
ds^2 = -\exp\left(-\frac{2m}{r}-2\right)dt^2 + \exp\left(\frac{2m}{r}+2\right)[dr^2 + r^2 d\theta^2 + r^2 sin^2\theta d\phi^2].
\end{equation}
Then, radius at the throat becomes $r=r_0=m$ (from Eq.(\ref{(43)})), for any positive values of $m$.
\\Similarly, from the Eq.(\ref{(73)}) , the NEC is violated near the throat region as,
\begin{equation}\label{(75)}
p_r(r_0)+\rho(r_0)=-\frac{e^{-4}}{2\pi m^2}<0.
\end{equation}
\\
\textbf{Case 2} If both $n=k=1$, the metric takes the form,
\begin{equation}\label{(77)}
ds^2 = -\exp\left(-\frac{2m}{r}-2\alpha r\right)dt^2 + \exp\left(\frac{2m}{r}+2\alpha r\right)[dr^2 + r^2 d\theta^2 + r^2 sin^2\theta d\phi^2].
\end{equation}
This metric is a special case of the metric Eq.(\ref{(14)}) or Eq.(\ref{(37)})(or the metric is same that expressed in Eq.(\ref{(47)})). Similarly the radius at the throat is again $r=r_0=m$, for $0<m<<\frac{1}{4\alpha}$. This metric again violates the NEC near the throat region.
\\
\textbf{Case 3} If both $n=k=-1$, the metric becomes,
\begin{equation}
ds^2 = -\exp\left(-\frac{2m}{r}-\frac{2}{\alpha r}\right)dt^2 + \exp\left(\frac{2m}{r}+\frac{2}{\alpha r}\right)[dr^2 + r^2 d\theta^2 + r^2 sin^2\theta d\phi^2].
\end{equation}
Again, the throat radius is similar with the metric Eq.(\ref{(50)}). We can also show that this metric violates NEC at the throat.
\\Now, let us consider an another exponential generalised metric as follows,
\begin{equation}\label{(78)}
ds^2 = -\exp\left(-\frac{2m}{r}-2\alpha^l r^l-2\alpha^q r^q\right)dt^2 + \exp\left(\frac{2m}{r}+2\alpha^n r^n+2\alpha^k r^k\right)[dr^2 + r^2 d\theta^2 + r^2 sin^2\theta d\phi^2],
\end{equation}
where $k$, $l$, $n$, $q$ are integers. The area of the spherical surface is same as that of the metric represented by Eq.(\ref{(52)}). From Einstein tensors and using the natural units, we will get
\begin{align}\label{(79)}
\begin{split}
p_r(r)={}&\frac{\exp\left(-\frac{2m}{r}-2r^k\alpha^k -2r^n\alpha^n\right)}{8\pi r^4}[-m^2+kr^{k+2}\alpha^k(2+kr^k\alpha^k)+2mr(lr^l\alpha^l+qr^q\alpha^q)\\&+r^2(-2lr^l\alpha^l(1+kr^k\alpha^k +nr^n\alpha^n)-2qr^q\alpha^q(1+kr^k\alpha^k+nr^n\alpha^n)+nr^n\alpha^n(2\\&+2kr^k\alpha^k + nr^n\alpha^n))],
\end{split}
\end{align}
\begin{align}\label{(80)}
\begin{split}
\rho(r)={}&-\frac{\exp\left(-\frac{2m}{r}-2r^k\alpha^k -2r^n\alpha^n\right)}{8\pi r^4}[m^2-2mr(kr^k\alpha^k +nr^n\alpha^n)+r^2(n^2r^{2n}\alpha^{2n}+2nr^n\alpha^n\\&(1+n+kr^k\alpha^k)+kr^k\alpha^k(2+k(2+r^k\alpha^k)))]
\end{split}
\end{align}
And
\begin{align}\label{(81)}
\begin{split}
p_r(r)+\rho(r)={}&\frac{\exp\left(-\frac{2m}{r}-2r^k\alpha^k -2r^n\alpha^n\right)}{4\pi r^4}[-m^2+kr^{k+1}(m-kr)\alpha^k +nr^{n+1}(m-nr)\alpha^n\\&-lr^{l+1}\alpha^l(-m+r(1+kr^k\alpha^k+nr^n\alpha^n))-qr^{q+1}\alpha^q(-m+r(1+kr^k\alpha^k \\&+nr^n\alpha^n))].
\end{split}
\end{align}
Now, let us consider some simple cases by taking some values of $n,k,l,q$.\\
\textbf{Case 1}If $l=q=n=k=0$, then the metric Eq.(\ref{(78)}) takes the form,
\begin{equation}\label{(82)}
ds^2 = -\exp\left(-\frac{2m}{r}-4\right)dt^2 + \exp\left(\frac{2m}{r}+4\right)[dr^2 + r^2 d\theta^2 + r^2 sin^2\theta d\phi^2].
\end{equation}
Then the radius at the throat comes out as $r=r_0=m$ from Eq.(\ref{(57)}). Again from Eq.(\ref{(81)}), at the throat we can see that,
\begin{equation}\label{(83)}
p_r(r_0)+\rho(r_0)=-\frac{e^{-6}}{4\pi m^2}<0.
\end{equation}
\\
\textbf{Case 2} If $l=q=0$ and $n=k=1$, the metric Eq.(\ref{(78)}) becomes,
\begin{equation}\label{(86)}
ds^2 = -\exp\left(-\frac{2m}{r}-4\right)dt^2 + \exp\left(\frac{2m}{r}+4\alpha r\right)[dr^2 + r^2 d\theta^2 + r^2 sin^2\theta d\phi^2].
\end{equation}
Then the radius at throat comes out as $r=\frac{-1+\sqrt{1+8\alpha m}}{4\alpha}$ and if $0<m<<\frac{1}{8\alpha}$, then the radius at the throat becomes $r=r_0=m$. Then from Eq.(\ref{(81)}), we will get,
\begin{equation}\label{(87)}
p_r(r_0)+\rho(r_0)=-\frac{e^{(-4\alpha m-2)}}{4\pi m^2}<0.
\end{equation}
\\
\textbf{Case 3} If $l=q=n=k=1$, then the metric Eq.(\ref{(78)}) takes the form,
\begin{equation}\label{(98)}
ds^2 = -\exp\left(-\frac{2m}{r}-4\alpha r\right)dt^2 + \exp\left(\frac{2m}{r}+4\alpha r\right)[dr^2 + r^2 d\theta^2 + r^2 sin^2\theta d\phi^2].
\end{equation}
Then the radius at the throat takes the form $r=\frac{-1+\sqrt{1+8\alpha m}}{4\alpha}$ and if $0<m<<\frac{1}{8\alpha}$, the radius at the throat becomes $r=r_0=m$. Again from Eq.(\ref{(81)}), we will get,
\begin{equation}\label{(99)}
p_r(r_0)+\rho(r_0)=-\frac{e^{(-4\alpha m-2)}}{4\pi m^2}[4\alpha^2 m^2+1]<0.
\end{equation}
\\
\textbf{Case 4} If $l=q=n=k=-1$, the metric takes the form,
\begin{equation}
ds^2 = -\exp\left(-\frac{2m}{r}-\frac{4}{\alpha r}\right)dt^2 + \exp\left(\frac{2m}{r}+\frac{4}{\alpha r}\right)[dr^2 + r^2 d\theta^2 + r^2 sin^2\theta d\phi^2],
\end{equation}
if $\alpha>>m$, the radius of the throat comes out as $r=r_0=m$ and the metric violates the NEC at the throat region.\\
All these metrics have some physical throat and also they violate the necessary Null Energy Condition(NEC). In this way we can check for different values of $l,q,n$ and $k$. So the metric represented by Eq.(\ref{(78)}) is a new traversable exponential wormhole metric, which doesnot fall into the Ellis Gibbons class \cite{martinez2021static}. In this metric the radius of the throat remains the same if we keep the values of $n$ and $k$ fixed and vary the values of $l$ and $q$. So all the metrics in which the values of $n$ and $k$ are same will exhibit the same type of throat. But the reverse is not true. Any change in the values of $n$ and $k$ will lead to a different throat. 
\\
Similarly we can show that
\begin{equation}\label{(100)}
ds^2 = -\exp\left(-\frac{2m}{r}-2\alpha^l r^l\right)dt^2 + \exp\left(\frac{2m}{r}+2\alpha^n r^n+2\alpha^k r^k\right)[dr^2 + r^2 d\theta^2 + r^2 sin^2\theta d\phi^2]
\end{equation}
and
\begin{equation}\label{(101)}
ds^2 = -\exp\left(-\frac{2m}{r}-2\alpha^l r^l-2\alpha^q r^q\right)dt^2 + \exp\left(\frac{2m}{r}+2\alpha^n r^n\right)[dr^2 + r^2 d\theta^2 + r^2 sin^2\theta d\phi^2]
\end{equation}
are also some traversable exponential wormhole metric. Where $l,q,n,k$ are integers.
\\
In this way, we can say that the following metric is a new class of traversable exponential wormhole metric.
\begin{align}\label{(102)}
\begin{split}
ds^2 = {}&-\exp\left(-\frac{2m}{r}-2\alpha^l r^l-2\alpha^q r^q.....-2\alpha^{\beta}r^{\beta}\right)dt^2 + 
\exp\left(\frac{2m}{r}+2\alpha^n r^n+2\alpha^k r^k+...+2\alpha^{\gamma}r^{\gamma}\right)\\&[dr^2 + r^2 d\theta^2 + r^2 sin^2\theta d\phi^2], 
\end{split}
\end{align}
where $l,q,\beta,n,k,\gamma$ etc are integers. By assigning different values to $l,q,\beta,n,k,\gamma$, we can form different exponential wormhole metric. Again for the same values of $n$, $k$....$\gamma$ the throat radius will be the same for same range of $m$.
\par More precisely we can write the above mentioned metrics as,
\begin{equation}\label{mg}
ds^2= -\exp\left(-\frac{2m}{r}-h(r)\right)dt^2+ \exp\left(\frac{2m}{r}+ j(r)\right)[dr^2+ r^2 d\theta^2 + r^2 sin^2\theta d\phi^2].
\end{equation}
Where $h(r)$ and $j(r)$ are functions of $r$, in which the powers of $r$ must be integer. The function $h(r)$ is related to redshift and the function $j(r)$ is related to the shape of the wormhole.
\par All these formulated exponetial metric clearly donot contain any kind of horizon, since $\forall r\in (0,+\infty)$, we have $g_{tt}\neq0$. We have calculated the location of the throat for all these metrics(for some specific ranges of $m$), where the area of the spherical surfaces are minimum and the flare out condition is satisfied. The power term $n,k,l,q.....$ etc should take only intergral values. 
\section{Stability}
In order to be stable, the wormhole should follow the flare-out condition. The flare out condition is more understandable through the embedding geometry. Now at $t=constant$ and $\theta=\frac{\pi}{2}$, the embedded spacetime of the metric Eq.(\ref{mg}) is given as
\begin{equation}\label{z}
ds_e^2=\exp\left(\frac{2m}{r}+ j(r)\right)[dr^2+ r^2 d\phi^2].
\end{equation}
In three dimensional Euclidean space the embedded surface has equation $z=z(r)$, so the metric of the surface can be written as,
\begin{equation}\label{z1}
ds^2_e=[1+(\frac{dz}{dr})^2]dr^2+e^{\frac{2m}{r}+j(r)}r^2 d\phi^2.
\end{equation}
Compairing the relations Eq.(\ref{z}) and Eq.(\ref{z1}), we get
\begin{equation}
\frac{dz}{dr}=\pm (e^{\frac{2m}{r}+j(r)}-1)^{\frac{1}{2}}.
\end{equation}
The flare-out condition is given by the minimality of the wormhole throat as,
\begin{equation}
\frac{d}{dz}(\frac{dr}{dz})>0
\end{equation}
Now, the flare out condition for the metrics Eq.(\ref{(2)}),(\ref{(14)}),(\ref{(22)}) and (\ref{(32)}) suggest that
\begin{equation}
m^2-r^2 \alpha >0
\end{equation}
i.e. the value of $\alpha$ should be very small for those temporal metrics.\\
For the metrics Eq.(\ref{(37)}) and (\ref{(70)}), from the flare out condition, we found that,
\begin{equation}
m-n r^{n+1} \alpha^n>0.
\end{equation}
Which suggests that the value of $\alpha$ should be small for the positive values of $n$ and $\alpha$ should be large for negative values of $n$. Similarly for the metrics Eq.(\ref{(52)}) and (\ref{(78)}), we get,
\begin{equation}
m-r(kr^k \alpha^k +n r^n \alpha^n)>0
\end{equation}
which also implies that $\alpha$ should be small for positive values of $n$, $k$ and $\alpha$ should be very large for the negative values of $n$, $k$.\\
Usually, the exoticity function $\zeta$ is used for the flare out condition, which is given as,
\begin{equation}
\zeta =\frac{\tau-\rho}{|\rho|}>0.
\end{equation}
where $\tau$ represents surface tension.
For all these constructed metrics, the value of $\zeta$ is positive for any values of $n$, $k$, $l$ etc. Which signifies that these wormholes obey flare-out conditions everywhere. It was assumed that wormhole should have a large surface tension compared to the energy density to continue the geometry. This condition seems to be physically logical. This violates the Weak Energy Condition to minimize the use of exotic matter.
\section{Curvature coordinate}
To go to "curvature coordinates", for the exponential metric Eq.(\ref{(2)}), Eq.(\ref{(14)}) and Eq.(\ref{(23)}) we make the coordinate transformation,
\begin{equation}\label{(103)}
r_s=r\exp\left(\frac{m}{r}+\alpha r\right) , dr_s=\exp\left(\frac{m}{r}+\alpha r\right)[1-\frac{m}{r}+\alpha r]dr.
\end{equation}
Now, the exponential metric Eq.(\ref{(2)}) in curvature coordinates becomes,
\begin{equation}\label{(104)}
ds^2=-\exp\left(-\frac{2m}{r}\right)dt^2+\frac{1}{(1-\frac{m}{r}+\alpha r)^2}dr_s^2+ r_s^2(d\theta^2+sin^2\theta d\phi^2).
\end{equation}
The metric Eq.(\ref{(14)}) takes the form,
\begin{equation}\label{(105)}
ds^2=-\exp\left(-\frac{2m}{r}-2\alpha^n r^n\right)dt^2+\frac{1}{(1-\frac{m}{r}+\alpha r)^2}dr_s^2+ r_s^2(d\theta^2+sin^2\theta d\phi^2).
\end{equation}
And the metric Eq.(\ref{(23)}) becomes,
\begin{equation}\label{(106)}
ds^2=-\exp\left(-\frac{2m}{r}-2\alpha^n r^n-2\alpha^l r^l\right)dt^2+\frac{1}{(1-\frac{m}{r}+\alpha r)^2}dr_s^2+ r_s^2(d\theta^2+sin^2\theta d\phi^2).
\end{equation}
Here $r$ is an implicit function of $r_s$. Now, for the minimum of the coordinate $r_s$
\begin{equation}\label{(108)}
\frac{dr_s}{dr}= \exp\left(\frac{m}{r}+\alpha r\right)\left[1-\frac{m}{r}+\alpha r\right] ; \frac{dr_s}{dr}=0.
\end{equation}
Therefore, $1-\frac{m}{r}+\alpha r=0$ and we will get $r=\frac{-1+\sqrt{1+4\alpha m}}{2\alpha}$
If $m<<\frac{1}{4\alpha}$, then $r=m$ and the minimum value of $r_s$ becomes
\begin{equation}\label{(109)}
r_s=m\exp\left(\alpha m+1\right).
\end{equation} 
So we have a stationary point at $r=m$,for $m<<\frac{1}{4\alpha}$, which corresponds to $r_s=m\exp(\alpha m+1)$
\begin{equation}\label{(110)}
\frac{d^2r_s}{dr^2}=\exp\left(\frac{m}{r}+\alpha r\right)\left[\frac{m^2}{r^3}-\frac{2m\alpha}{r}+\alpha^2 r+2\alpha +1\right].
\end{equation}
At $r=m$
\begin{equation}\label{(111)}
\frac{d^2r_s}{dr^2}|_{r=m}=\exp\left(\alpha m+1\right)\left[\alpha^2 m+\frac{1}{m}+1\right]>0.
\end{equation}
The curvature coordinate $r_s$ therefore has a minimum at $r_s=m\exp\left(\alpha m+1\right)$, and in these curvature coordinates the exponential metrics exhibit a wormhole throat at $r_s=m\exp\left(\alpha m+1\right)$.
\\
\\
Now, for the metric Eq.(\ref{(37)}) and Eq.(\ref{(70)}), let us consider the coordinate transformation,
\begin{equation}\label{(112)}
r_s=r\exp\left(\frac{m}{r}+\alpha^n r^n\right); dr_s=\exp\left(\frac{m}{r}+\alpha^n r^n\right)\left[1-\frac{m}{r}+n\alpha^n r^n\right]dr.
\end{equation}
So, the metric Eq.(\ref{(37)}) and Eq.(\ref{(70)}) takes the form,
\begin{equation}\label{(113)}
ds^2=-\exp\left(-\frac{2m}{r}-2\alpha r\right)dt^2+\frac{1}{(1-\frac{m}{r}+n\alpha^n r^n)^2}dr_s^2+ r_s^2(d\theta^2+sin^2\theta d\phi^2)
\end{equation}
and
\begin{equation}\label{(114)}
ds^2=-\exp\left(-\frac{2m}{r}-2\alpha^k r^k\right)dt^2+\frac{1}{(1-\frac{m}{r}+n\alpha^n r^n)^2}dr_s^2+ r_s^2(d\theta^2+sin^2\theta d\phi^2),
\end{equation}
respectively.\\
Here, again $r$ is an implicit function of $r_s$. The minimum of the coordinate $r_s$ is,
\begin{equation}\label{(115)}
\begin{multlined}
\frac{dr_s}{dr}=\exp\left(\frac{m}{r}+\alpha^n r^n\right)\left[1-\frac{m}{r}+n\alpha^n r^n\right]; \frac{dr_s}{dr}=0.
\end{multlined}
\end{equation}
Therefore $1-\frac{m}{r}+n\alpha^n r^n=0$ and by solving this equation for different values of $n$, we will get the radius at the throat. Again by substituting this value of $r$ in Eq.(\ref{(112)}), we will get the minimum value of $r_s$. The obtained value of $r_s$ will be the radius of the exponential metrics Eq.(\ref{(37)}) and Eq.(\ref{(70)}) in curvature coordinates.
\\
\\
Again, for Eq.(\ref{(52)}) and Eq.(\ref{(78)}), let us consider the transformation as,
\begin{equation}\label{(116)}
r_s=r\exp\left(\frac{m}{r}+\alpha^n r^n+\alpha^k r^k\right); dr_s=\exp\left(\frac{m}{r}+\alpha^n r^n+\alpha^k r^k\right)[1-\frac{m}{r}+n\alpha^n r^n+k\alpha^k r^k]dr.
\end{equation}
The metric Eq.(\ref{(52)}) and Eq.(\ref{(78)}) take the form as,
\begin{equation}\label{(117)}
ds^2=-\exp\left(-\frac{2m}{r}-2\alpha r\right)dt^2+\frac{1}{(1-\frac{m}{r}+n\alpha^n r^n+k\alpha^k r^k)^2}dr_s^2+ r_s^2(d\theta^2+sin^2\theta d\phi^2)
\end{equation}
and
\begin{equation}\label{(118)}
ds^2=-\exp\left(-\frac{2m}{r}-2\alpha^l r^l-2\alpha^q r^q\right)dt^2+\frac{1}{(1-\frac{m}{r}+n\alpha^n r^n+k\alpha^k r^k)^2}dr_s^2+ r_s^2(d\theta^2+sin^2\theta d\phi^2),
\end{equation}
respectively.\\
Again for the minimum value of the coordinate $r_s$,
\begin{equation}\label{(119)}
 \frac{dr_s}{dr}=\exp\left(\frac{m}{r}+\alpha^n r^n+\alpha^k r^k\right)\left[1-\frac{m}{r}+n\alpha^n r^n+k\alpha^k r^k\right]; \frac{dr_s}{dr}=0.
\end{equation}
Therefore $1-\frac{m}{r}+n\alpha^n r^n+k\alpha^k r^k=0$, similarly by solving this equation we will get the minimum value of $r$. Again by substituting this value of $r$ in Eq.(\ref{(116)}), we will get the minimum value of $r_s$, which can be considered as the radius of throat of the exponential metrics Eq.(\ref{(52)}) and Eq.(\ref{(78)}) in curvature coordinates.
\\
\section{Curvature Tensor and Singularity}
The non-zero components of Riemann tensor, Ricci tensor, Kretschmann constants for the metric Eq.(\ref{(2)}) are
\begin{equation}\label{(125)}
R=-\frac{2e^{(-\frac{2m}{r}-2\alpha r})}{r^4}[m^2-mr^2 \alpha+4r^3 \alpha+r^4 \alpha^2],
\end{equation}
\begin{align}\label{(126)}
\begin{split}
R_{abcd}R^{abcd}={}&\frac{4e^{(-\frac{4m}{r}-4\alpha r)}}{r^8}[7m^4-4m^3 r(4+3r\alpha)-4mr^4 \alpha(1+r \alpha(3+r\alpha))+r^6 \alpha^2(6+\\
&r \alpha(4+r \alpha))+m^2 r^2(12+r \alpha(20+9r \alpha))],
\end{split}
\end{align}
\begin{equation}\label{(127)}
C_{abcd}C^{abcd}=4e^{(-\frac{4m}{r}-4\alpha r)}\frac{(4m^2+r^3 \alpha(1+r \alpha)-2mr(3+2r \alpha))^2}{3r^8}.
\end{equation}

\begin{align}
\begin{split}
R_{ab}R^{ab}={}& \frac{e^{-\frac{4m}{r}-4\alpha r}}{r^8}[4m^4-4m^3 r^2 \alpha+4m^2 r^3 \alpha(2+r \alpha)-4m r^5 \alpha^2(4+r \alpha)+2r^6 \alpha^2(11+\\
&r \alpha(6+r \alpha))]
\end{split}
\end{align}
The non zero Electric parts of the Weyl tensor are
\begin{align}
\begin{split}
{}&E_{rr}=-2E_{\theta \theta}=-2E_{\phi \phi}=\frac{e^{-\frac{2m}{r}}}{3r^4}[4m^2+r^4 \alpha^2-4mr(1+r \alpha)];\\
&E_{tt}=\frac{e^{-\frac{6m}{r}-2r\alpha}}{3r^4}[m^2+r^4 \alpha^2+mr(2-r\alpha)]
\end{split}
\end{align}
and
\begin{align}
\begin{split}
E_{ab}E^{ab}={}&\frac{e^{-\frac{8m}{r}-4r \alpha}}{36r^{12}}[4m^4 (4+17 r^4)+ r^8 (1+8 r^4)\alpha^4-8m r^5  \alpha^2 (1+2 r^4 +r \alpha+5 r^5 \alpha)-\\
&8m^3 r(4+14 r^4 +4r \alpha+17 r^5 \alpha)+4 m^2  r^2 (4+r(20 r^3 +4(2+7 r^4 )\alpha+3r(2+\\
&9 r^4 ) \alpha^2 ))+(4m^2 + r^4 \alpha^2-4mr(1+r \alpha))^2 cosec^4 \theta]
\end{split}
\end{align}
All these components are finite and they donot diverge at $r=0$ and $r=r_0=m$. So, we can say that the metric Eq.(\ref{(2)}) doesnot contain any kind of Weyl and oscillating Ricci singularity \cite{king1974new}. Now, in order to inspect the parallelly propagated(p.p.) curvature singularity, we need to construct a p.p. basis along the relevant curves. For null geodesics with tangent vector $h^a$, the p.p. pseudo-orthonormal basis is \cite{harada2022complete},
\begin{equation}
h_a=e^{-\frac{\alpha r}{2}}\{\frac{e^{-\frac{m}{r}}}{\sqrt{2}},0,0,\pm \frac{e^{\frac{m}{r}+ \alpha r}}{\sqrt{2}}\}; m_a=e^{-\frac{\alpha r}{2}}\{\frac{e^{-\frac{m}{r}+\alpha r}}{\sqrt{2}},0,0,\mp\frac{e^{\frac{m}{r}+2 \alpha r}}{\sqrt{2}}\}
\end{equation}
These satisfy,
\begin{equation}
h^a h_a=0; m^a m_a=0; h^a m_a=-1; h^a e_{(A)a}=0; m^a e_{(A)a}=0; g^{ab} e_{(A)a} e_{(B)a}=\delta_{AB}
\end{equation}
where $A$ and $B$ run over $2$ and $3$ and
\begin{align}
\begin{split}
{}& e_{(0)a}=\{-e^{-\frac{m}{r}},0,0,0\}; e_{(1)a}=\{0,e^{\frac{m}{r}+ \alpha r},0,0\}\\
& e_{(2)a}=\{0,0,r e^{\frac{m}{r}+ \alpha r},0\}; e_{(3)a}=\{0,0,0,r sin\theta e^{\frac{m}{r}+ \alpha r}\}
\end{split}
\end{align}
On these basis we could find that $R_{abcd}h^a m^b h^c m^d$, $R_{abcd}h^a h^b$, $R_{abcd}m^a m^b$ and $R_{abcd}h^a m^b$ show finite value everywhere on the space-time. So we can say that the metric doesnot contain any kind of p.p. curvature singularity.
\\
Again, for the metric Eq.(\ref{(23)}),
\begin{align}\label{(147)}
\begin{split}
R={}&\frac{2 e^{-\frac{2m}{r}-2\alpha r}}{r^4}[-m^2+mr(\alpha r+l r^l \alpha^l+n r^n \alpha^n)+r^2(-4 \alpha r- \alpha^2 r^2+l r^l \alpha^l+l^2 r^l \alpha^l-\\
&l^2 r^{2l} \alpha^{2l}+l r^{l+1} \alpha^{l+1}-n^2 r^{2n} \alpha^{2n} +n r^n \alpha^n (1+n+\alpha r -2l r^l \alpha^l))],
\end{split}
\end{align}

\begin{align}\label{(149)}
\begin{split}
R_{abcd}R^{abcd}={}&\frac{4e^{(-\frac{4m}{r}-4\alpha r)}}{r^8}[7m^4-4m^3 r(4+3 \alpha r+4l r^l \alpha^l+4n r^n \alpha^n)+m^2 r^2(12+20 \alpha r+\\
&9 r^2 \alpha^2+24l r^l \alpha^l-4l^2 r^l \alpha^l+15l^2 r^{2l} \alpha^{2l}+18l r^{l+1} \alpha^{l+1}+15 n^2 r^{2n} \alpha^{2n}+2nr^n \alpha^n\\
&(12-2n+9 \alpha r+ 15l r^l \alpha^l))+2mr^3(-2\alpha r-6 r^2 \alpha^2-2 r^3 \alpha^3-4lr^l \alpha^l+2l^2 r^l \alpha^l\\
&-7l^2 r^{2l}\alpha^{2l}+3l^3 r^{2l} \alpha^{2l}-3l^3 r^{3l} \alpha^{3l}-7lr^{l+1} \alpha^{l+1}+l^2 r^{l+1} \alpha^{l+1}-3lr^{l+2} \alpha^{l+2}-\\
&6l^2 r^{2l+1} \alpha^{2l+1}-3n^3 r^{3n} \alpha^{3n}+n^2 r^{2n} \alpha^{2n}(-7+3n-6\alpha r-9lr^l \alpha^l)-nr^n \alpha^n(-4\\
&+2n-7 \alpha r+n \alpha r-3r^2 \alpha^2-9l^2 r^{2l} \alpha^{2l}+lr^l \alpha^l(-14+3l+3n-12 \alpha r)))+r^4\\
&(l^4 r^{4l} \alpha^{4l}+n^4 r^{4n} \alpha^{4n}+2l^3 r^{3l} \alpha^{3l}(1-l+\alpha r)+2n^3 r^{3n} \alpha^{3n}(1-n+\alpha r +2l r^l \alpha^l)\\
&+n^2 r^{2n} \alpha^{2n}(3-2n+n^2+6 \alpha r-2n\alpha r+3r^2 \alpha^2 +6l^2 r^{2l} \alpha^{2l}-2lr^l \alpha^l(-3+l\\
&+2n-3 \alpha r))+l^2 r^{2l} \alpha^{2l}(l^2-2l(1+\alpha r)+3(1+ \alpha r)^2)+r^2 \alpha^2(6+\alpha r(4+ \alpha r))\\
&+2lnr^{n+l} \alpha^{n+l}(-(-3+n-3\alpha r)(1+\alpha r)+2l^2 r^l \alpha^l(-1+r^l \alpha^l)+l(-1+n-\\
&\alpha r+r^l \alpha^l(3-n+3 \alpha r))))],
\end{split}
\end{align}
\begin{align}\label{(150)}
\begin{split}
C_{abcd}C^{abcd}={}&\frac{4e^{(-\frac{4m}{r}-4\alpha r)}}{3r^8}[4m^2-6mr-4mr^2 \alpha+ \alpha r^3+\alpha^2 r^4-4lmr^{l+1}+2lr^{l+2} \alpha^l-l^2 r^{l+2} \alpha^l\\
&+l^2 r^{2l+2} \alpha^{2l}+2l r^{l+3} \alpha^{l+1}-4mnr^{n+1} \alpha^n+2nr^{n+2} \alpha^n-n^2 r^{n+2} \alpha^n+n^2 r^{2n+2} \alpha^{2n}\\
&+2nr^{n+3} \alpha^{n+1}+2lnr^{n+l+2} \alpha^{n+l}]^2,
\end{split}
\end{align}
Again the non zero electric components of Weyl tensor are,
\begin{align}
\begin{split}
{}&E_{rr}=-2E_{\theta \theta}=-2E_{\phi \phi}=\frac{e^{-\frac{2m}{r}-2r^l \alpha^l -2r^n \alpha^n}}{3r^4}[4m^2-4mr(1+\alpha r+ lr^l \alpha^l+nr^n \alpha^n)+r^2(r^2 \alpha^2\\
&+lr^l \alpha^l-l^2 r^l \alpha^l+l^2 r^{2l} \alpha^{2l}+2l r^{l+1} \alpha^{l+1}+n^2 r^{2n} \alpha^{2n} +nr^n \alpha^n(1-n+2 \alpha r +2lr^l \alpha^l))]
\end{split}
\end{align}
and
\begin{align}
\begin{split}
E_{tt}={}&\frac{e^{-\frac{6m}{r}-2\alpha r-4r^l \alpha^l-4r^n \alpha^n}}{3r^4}[-m^2+mr(-2+\alpha r+lr^l \alpha^l+nr^n \alpha^n)+r^2(-r^2 \alpha^2-lr^l \alpha^l+\\
&l^2 r^l \alpha^l+lr^{l+1} \alpha^{l+1}-n^2 r^{2n} \alpha^{2n}+nr^n \alpha^n(-1+n+\alpha r-2lr^l \alpha^l))]
\end{split}
\end{align}
Similarly we can also show that $R_{ab}R^{ab}$ and $E_{ab}E^{ab}$ have finite values and donot diverge as $r\to 0$ and $r\to r_0$. That is the metric doesnot contain any kind of singularity. The p.p. pseudo orthonormal basis for this metric is
\begin{align}
\begin{split}
{}& h_a= e^{-\frac{\alpha r}{2}+\frac{\alpha^l r^l}{2}+\frac{\alpha^n r^n}{2}}\{\frac{e^{-\frac{m}{r}-\alpha^n r^n-\alpha^l r^l}}{\sqrt{2}},0,0,\pm \frac{e^{\frac{m}{r}+\alpha r}}{\sqrt{2}}\};\\
& m_a= e^{-\frac{\alpha r}{2}+\frac{\alpha^l r^l}{2}+\frac{\alpha^n r^n}{2}}\{\frac{e^{-\frac{m}{r}+\alpha r-2 \alpha^l r^l-2 \alpha^n r^n}}{\sqrt{2}},0,0,\mp \frac{e^{\frac{m}{r}+2 \alpha r-\alpha^n r^n-\alpha^l r^l}}{\sqrt{2}}\}
\end{split}
\end{align}
On these basis the value of $R_{abcd}h^a m^b h^c m^d$, $R_{abcd}h^a h^b$, $R_{abcd}m^a m^b$ and $R_{abcd}h^a m^b$ all remain finite everywhere. So this metric is free from any kind of p.p. curvature singularities.
\par Again for the metric Eq.(\ref{(52)}),
\begin{align}\label{(164)}
\begin{split}
R={}&-\frac{2 e^{-\frac{2m}{r}-2r^k \alpha^k-2r^n \alpha^n}}{r^4}[m^2-mr(\alpha r+kr^k \alpha^k +nr^n \alpha^n)+r^2(-2 \alpha r+ \alpha^2 r^2+2k r^k \alpha^k+\\
& 2k^2 r^k \alpha^k +k^2 r^{2k} \alpha^{2k}- kr^{k+1} \alpha^{k+1} + n^2 r^{2n} \alpha^{2n} +nr^n \alpha^n(2+2n- \alpha r+ 2k r^k \alpha^k))],
\end{split}
\end{align}
and 
\begin{align}\label{(165)}
\begin{split}
R_{abcd}R^{abcd}={}&\frac{4 e^{-\frac{4m}{r}-4r^k \alpha^k -4r^n \alpha^n}}{r^8}[7m^4-4m^3 r(4+ 4 \alpha r+ 3k r^k \alpha^k + 3nr^n \alpha^n)+m^2 r^2(12+\\
& 20 \alpha r+ 15 \alpha^2 r^2+ 20k r^k \alpha^k+ 9k^2 r^{2k} \alpha^{2k}+ 18k r^{k+1} \alpha^{k+1}+ 9n^2 r^{2n} \alpha^{2n}+ 2nr^n \alpha^n \\
&(10+9\alpha r + 9kr^k \alpha^k))+r^4(2r^2 \alpha^2+ r^4 \alpha^4+4k^2 r^{2k} \alpha^{2k}+ 2k^4 r^{2k} \alpha^{2k}+4k^3 r^{3k} \alpha^{3k}\\
&+ k^4 r^{4k} \alpha^{4k} +4k r^{k+2} \alpha^{k+2}+ 2kr^{k+3} \alpha^{k+3}+3k^2 r^{2k+2} \alpha^{2k+2} + n^4 r^{4n} \alpha^{4n}+ 4n^3 r^{3n} \alpha^{3n}\\
&(1+k r^k \alpha^k)+n^2 r^{2n} \alpha^{2n} (4+2n^2+ 3r^2 \alpha^2+ 12k r^k \alpha^k+ 6k^2 r^{2k} \alpha^{2k})+2nr^n \alpha^n\\
&(6k^2 r^{2k} \alpha^{2k}+2k^3 r^{3k} \alpha^{3k} +r^2 \alpha^2(2+ \alpha r)+kr^k \alpha^k(4+2kn+3 r^2 \alpha^2)))+2mr^3\\
&(-2 \alpha r-4r^2 \alpha^2- 3r^3 \alpha^3- 4k r^k \alpha^k+2k^2 r^k \alpha^k- 6k^2 r^{2k} \alpha^{2k}-2k^3 r^{3k} \alpha^{3k}-\\
&6kr^{k+1} \alpha^{k+1}-6k r^{k+2} \alpha^{k+2}-3k^2 r^{2k+1} \alpha^{2k+1}-2n^3 r^{3n} \alpha^{3n}-3n^2 r^{2n} \alpha^{2n}(2+ \alpha r\\
&+ 2k r^k \alpha^k)-2n r^n \alpha^n(2-n+3(k^2 r^{2k} \alpha^{2k} + \alpha r(1+ \alpha r)+k r^k \alpha^k(2+ \alpha r))))],
\end{split}
\end{align}

\begin{align}
\begin{split}
C_{abcd}C^{abcd}={}&\frac{4 e^{-\frac{4m}{r}-4r^k \alpha^k- 4r^n \alpha^n}}{3r^8}[4m^2-2mr(3+2 \alpha r+2k r^k \alpha^k+2n r^n \alpha^n)+r^2(\alpha r+\alpha^2 r^2\\
&+ 2kr^k \alpha^k- k^2 r^k \alpha^k+ k^2 r^{2k} \alpha^{2k}+ 2kr^{k+1} \alpha^{k+1}+ n^2 r^{2n} \alpha^{2n}+ nr^n \alpha^n(2-n+\\
& 2 \alpha r+2kr^k  \alpha^k))]^2.
\end{split}
\end{align}
Again the non-zero Electric components of Weyl tensor are,
\begin{align}
\begin{split}
E_{rr}=-2E_{\theta \theta}=-2E_{\phi\phi}={}&\frac{e^{-\frac{2m}{r}-2 \alpha r}}{3r^4}[4m^2-4mr(1+\alpha r+k r^k \alpha^k+n r^n \alpha^n)+r^2(\alpha^2 r^2+\\
&k^2 r^{2k} \alpha^{2k}+nr^n \alpha^n-n^2 r^n \alpha^n+n^2 r^{2n} \alpha^{2n}+2nr^{n+1} \alpha^{n+1}+kr^k \alpha^k\\
&(1-k+2 \alpha r+ 2nr^n \alpha^n))]
\end{split}
\end{align}
and
\begin{align}
\begin{split}
E_{tt}={}&\frac{e^{-\frac{6m}{r}-4 \alpha r-2r^k \alpha^k-2r^n \alpha^n}}{3r^4}[-m^2+mr(-2+ \alpha r+kr^k \alpha^k+nr^n \alpha^n)+r^2(-r^2 \alpha^2-k^2 r^{2k} \alpha^{2k}\\
&+2n r^n \alpha^n-2n^2 r^n \alpha^n-n^2 r^{2n} \alpha^{2n}+nr^{n+1} \alpha^{n+1}+kr^k \alpha^k(2-2k+ \alpha r-2nr^n \alpha^n))].
\end{split}
\end{align}
We can also find that $R_{ab}R^{ab}$ and $E_{ab}E^{ab}$ are finite as $r\to 0$ and $r\to r_0=m$. The p.p. pseudo orthonormal basis for this metric is,
\begin{align}
\begin{split}
{}& h_a=e^{\frac{\alpha r}{2}-\frac{\alpha^n r^n}{2}-\frac{\alpha^k r^k}{2}}\{\frac{e^{-\frac{m}{r}-\alpha r}}{\sqrt{2}},0,0,\pm \frac{e^{\frac{m}{r}+\alpha^n r^n+\alpha^k r^k}}{\sqrt{2}}\};\\
&m_a=e^{\frac{\alpha r}{2}-\frac{\alpha^n r^n}{2}-\frac{\alpha^k r^k}{2}}\{\frac{e^{-\frac{m}{r}-2\alpha r+\alpha^n r^n+\alpha^k r^k}}{\sqrt{2}},0,0,\mp \frac{e^{\frac{m}{r}-\alpha r+2 \alpha^n r^n+ 2 \alpha^k r^k}}{\sqrt{2}}\}.
\end{split}
\end{align}
The value of $R_{abcd}h^a m^b h^c m^d$, $R_{abcd}h^a h^b$, $R_{abcd}m^a m^b$ and $R_{abcd}h^a m^b$ again remain finite on these basis. That is the spacetime is free from p.p. curvature singularity.
\par Lastly, for the metric Eq.(\ref{(78)}),
\begin{align}\label{(168)}
\begin{split}
R={}&\frac{2e^{(-\frac{2m}{r}-2r^k\alpha^k-2r^n\alpha^n)}}{r^4}[-m^2+mr(kr^k\alpha^k+lr^l\alpha^l+nr^n\alpha^n+qr^q\alpha^q)+r^2(-q^2r^{2q}\alpha^{2q}-\\
&k^2r^k\alpha^k(2+r^k\alpha^k)-l^2r^l\alpha^l(-1+r^l\alpha^l)-nr^n\alpha^n(2+2n+nr^n\alpha^n)+qr^q\alpha^q(1+q+\\
&nr^n\alpha^n)+lr^l\alpha^l(1+nr^n\alpha^n-2qr^q\alpha^q)+kr^k\alpha^k(-2+lr^l\alpha^l-2nr^n\alpha^n+qr^q\alpha^q))]
\end{split}
\end{align}
and
\begin{align}\label{(169)}
\begin{split}
C_{abcd}C^{abcd}={}&\frac{4e^{-\frac{4m}{r}-4r^k \alpha^k-4r^n \alpha^n}}{3r^8}[4m^2+r^2(q^2 r^{2q} \alpha^{2q}+ k^2 r^k \alpha^k(-1+r^k \alpha^k)+l^2 r^l \alpha^l(-1+r^l \alpha^l)\\
&-q r^q \alpha^q(-2+q-2nr^n \alpha^n)+nr^n \alpha^n(2-n+nr^n \alpha^n)+2lr^l \alpha^l(1+nr^n \alpha^n+\\
&q r^q \alpha^q)+2kr^k \alpha^k(1+lr^l \alpha^l+nr^n \alpha^n+ qr^q\alpha^q))-2mr(3+2(kr^k \alpha^k+lr^l \alpha^l+\\
&nr^n \alpha^n+qr^q \alpha^q))]^2
\end{split}
\end{align}
The non zero Electric components of the Weyl tensor are,
\begin{align}
\begin{split}
E_{rr}=-2E_{\theta \theta}=-2E_{\phi \phi}={}&\frac{e^{-\frac{2m}{r}-2r^q \alpha^q-2r^s \alpha^s}}{3r^4}[4m^2-4mr(1+nr^n \alpha^n+kr^k \alpha^k +qr^q \alpha^q +lr^l \alpha^l)\\
&+r^2(r^{2l} l^2 \alpha^{2l}+n^2 r^n \alpha^n(-1+r^n \alpha^n)+k^2 r^k \alpha^k(-1+r^k \alpha^k)-lr^l \alpha^l\\
&(-1+l-2q r^q \alpha^q)+ qr^q \alpha^q(1-q+q r^q \alpha^q)+k r^k \alpha^k(1+2qr^q \alpha^q+ \\
&2r^l l \alpha^l)+nr^n \alpha^n(1+2k r^k \alpha^k+ 2qr^q \alpha^q + 2lr^l \alpha^l))],
\end{split}
\end{align}
and
\begin{align}
\begin{split}
E_{tt}={}&\frac{e^{-\frac{6m}{r}-2r^n \alpha^n-2r^k \alpha^k-4r^q \alpha^q-4r^l \alpha^l}}{3r^4}[-m^2+mr(-2+nr^n \alpha^n+kr^k \alpha^k+ qr^q \alpha^q+ lr^l \alpha^l)+r^2\\
&(-r^{2l} l^2 \alpha^{2l}- n^2 r^n \alpha^n(2+r^n \alpha^n)-k^2 r^k \alpha^k(2+r^k \alpha^k)+lr^l \alpha^l(-1+l-2qr^q \alpha^q)-qr^q \alpha^q\\
&(1-q+qr^q \alpha^q)+kr^k \alpha^k(2+qr^q \alpha^q+lr^l \alpha^l)+nr^n \alpha^n(2-2kr^k \alpha^k+qr^q \alpha^q+ lr^l \alpha^l))]
\end{split}
\end{align}
All these components (including $R_{ab}R^{ab},E_{ab}E^{ab}$) are finite at everypoint. To check the p.p. curvature singularity, let us construct a p.p. pseudo orthonormal basis,
\begin{align}
\begin{split}
{}& h_a=e^{\frac{1}{2}(\alpha^l r^l+\alpha^q r^q-\alpha^n r^n-\alpha^k r^k)}\{\frac{e^{-\frac{m}{r}-\alpha^l r^l-\alpha^q r^q}}{\sqrt{2}},0,0,\pm \frac{e^{\frac{m}{r}+\alpha^n r^n+\alpha^k r^k}}{\sqrt{2}}\};\\
& m_a=e^{\frac{1}{2}(\alpha^l r^l+\alpha^q r^q-\alpha^n r^n-\alpha^k r^k)}\{\frac{e^{-\frac{m}{r}+\alpha^n r^n+\alpha^k r^k-2\alpha^l r^l-2\alpha^q r^q}}{\sqrt{2}},0,0\mp \frac{e^{\frac{m}{r}+2\alpha^n r^n+2\alpha^k r^k-\alpha^l r^l-\alpha^q r^q}}{\sqrt{2}}\}
\end{split}
\end{align}
Here we can again see that the values of $R_{abcd} h^a m^b h^c m^d$, $R_{abcd} h^a h^b$, $R_{abcd} m^a m^b$ and $R_{abcd} h^a m^b$ remain finite at each and every point suggesting that the metric is free from any kind of p.p curvature singularities.
\par All the curvature components and scalar invariants of the metrics Eq.(\ref{(2)}), Eq.(\ref{(14)}), Eq.(\ref{(23)}), Eq.(\ref{(37)}), Eq.(\ref{(52)}), Eq.(\ref{(70)}) and Eq.(\ref{(78)}) are finite everywhere and interestingly they are finite at the throat and take maximum values near the throat region. Which implies that all these metrics donot contain any types of singularity. All these components decay to zero both as $r\to\infty$ and as $r\to0$.
  \par Also near the throat region we can see that for all timelike, null or spacelike vectors $l^a$ one has
\begin{equation}
R_{ab}l^al^b \leq 0.
\end{equation}
That is, near the throat region, all these metrics violate the null Ricci convergence condition \cite{martin2017classical, visser1995lorentzian} which is necessary for understanding the flare out at the throat of the traversable wormhole.

\section{Effective refractive index}

\par Our first modified metric Eq.(\ref{(2)}) can be written in the form
\begin{equation}
ds^2= e^{\frac{2m}{r}+2\alpha r}[-e^{-\frac{4m}{r}-2\alpha r}dt^2+(dr^2+r^2(d\theta^2+sin^2\theta d\phi^2))].
\end{equation}
Since the overall conformalfactor is irrelevant for photon propagation, so we can consider,
\begin{equation}
ds^2=-e^{-\frac{4m}{r}-2\alpha r}dt^2+(dr^2+r^2(d\theta^2+sin^2\theta d\phi^2)).
\end{equation} 
Which is
\begin{equation}
ds^2=-e^{-\frac{4m}{r}-2\alpha r}dt^2+(dx^2 +dy^2+dz^2).
\end{equation}
This metric corresponds to a coordinate sped of light $c(r)=e^{-\frac{2m}{r}-\alpha r}$ or equivalently an effective local refractive index
\begin{equation}
n_1(r)=\frac{1}{v_p}=e^{\frac{2m}{r}+\alpha r}.
\end{equation}
Where $v_p$ denotes the phase velocity of a photon as measured by a distant observer. In general for a space-time in isotropic coordinates (Eq.(\ref{isotropic})) the refractive index $n(r)$ with respect to vacuum can be written as,
\begin{equation}
n(r)=\frac{1}{v_p}=\sqrt{\frac{b(r)}{a(r)}}.
\end{equation}
We can rearrange the above equation as,
\begin{equation}
1=v_p \sqrt{\frac{b(r)}{a(r)}}.
\end{equation}
Which implies that at any given position, the speed of light is universal in all inertial frames of reference.
\\Let us consider the Schwarzschild spacetime in isotropic coordinates.
\begin{equation}
ds^2_{Sch}=-{\left(\frac{1-\frac{m}{2r}}{1+\frac{m}{2r}}\right)}^2dt^2+{\left(1+\frac{m}{2r}\right)}^4{dr^2+r^2(d\theta^2+sin^2\theta d\phi^2)}.
\end{equation} 
Now, the effective refractive index of Schwarzschild spacetime in isotropic coordinates is \cite{delphenich2020geodesics},
\begin{equation}\label{sch}
n_{Sch}(r)=\frac{(1+\frac{m}{2r})^3}{|1-\frac{m}{2r}|},
\end{equation}

\begin{figure}[H]
\begin{subfigure}[b]{0.4\textwidth}
\includegraphics[width=\textwidth]{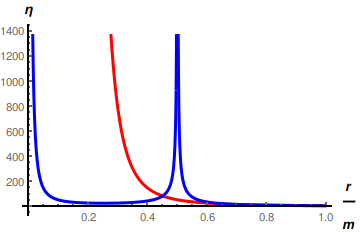}
\caption{\fcolorbox{red}{red}{\rule{0pt}{2.5pt}\rule{0pt}{2.5pt}} exponetial metric, \fcolorbox{blue}{blue}{\rule{0pt}{2.5pt}\rule{0pt}{2.5pt}} Schwarzschild}
\end{subfigure}
\hfill
\begin{subfigure}[b]{0.4\textwidth}
\includegraphics[width=\textwidth]{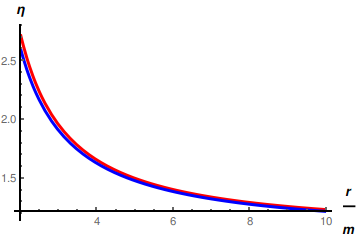} 
\caption{\fcolorbox{red}{red}{\rule{0pt}{2.5pt}\rule{0pt}{2.5pt}} exponential metric, \fcolorbox{blue}{blue}{\rule{0pt}{2.5pt}\rule{0pt}{2.5pt}} Schwarzschild} 
\end{subfigure}
\caption{The graph shows the refractive index for the exponential metric Eq.(\ref{(2)}) compared to the Schwarzschild metric in the isotropic coordinate. The left plot is for strong field region $r\approx\frac{m}{2}$ and the right plot is for relatively small $r\geq2m$.}
\label{FIG.1}
\end{figure}

Similarly, for the metric Eq.(\ref{(23)}) and (\ref{(52)}), the effective refractive index comes out as,
\begin{equation}
n_2(r)=\exp(\frac{2m}{r}+\alpha^n r^n+\alpha^lr^l+\alpha r); n_2(r)=\exp(\frac{2m}{r}+\alpha^n r^n+\alpha^kr^k+\alpha r).
\end{equation} 
From this we can consider some special cases,
\\if $n=l=0$ or $n=k=0$ then $n_{21}(r)=\exp(\frac{2m}{r}+\alpha r+2)$,
\\if $n=l=-1$ or $n=k=-1$, then $n_{22}(r)=\exp(\frac{2m}{r}+\alpha r + \frac{2}{\alpha r})$,
\\if $n=l=1$ or $n=k=1$ then $n_{23}(r)=\exp(\frac{2m}{r}+3\alpha r)$.
\begin{figure}[H]
\begin{subfigure}[b]{0.4\textwidth}
\includegraphics[width=\textwidth]{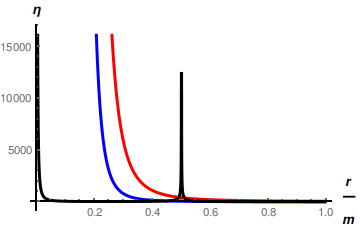}
\caption{\fcolorbox{red}{red}{\rule{0pt}{2.5pt}\rule{0pt}{2.5pt}} $n_{21}$, \fcolorbox{black}{black}{\rule{0pt}{2.5pt}\rule{0pt}{2.5pt}} $n_{22}$, \fcolorbox{green}{green}{\rule{0pt}{2.5pt}\rule{0pt}{2.5pt}} $n_{23}$, \fcolorbox{blue}{blue}{\rule{0pt}{2.5pt}\rule{0pt}{2.5pt}} $n_{Sch}$}
\end{subfigure}
\hfill
\begin{subfigure}[b]{0.4\textwidth}
\includegraphics[width=\textwidth]{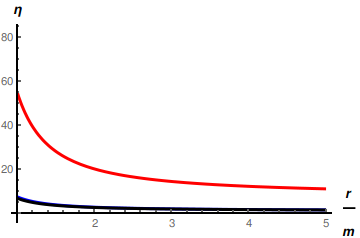} 
\caption{\fcolorbox{red}{red}{\rule{0pt}{2.5pt}\rule{0pt}{2.5pt}} $n_{21}$, \fcolorbox{black}{black}{\rule{0pt}{2.5pt}\rule{0pt}{2.5pt}} $n_{22}$, \fcolorbox{green}{green}{\rule{0pt}{2.5pt}\rule{0pt}{2.5pt}} $n_{23}$, \fcolorbox{blue}{blue}{\rule{0pt}{2.5pt}\rule{0pt}{2.5pt}} $n_{Sch}$} 
\end{subfigure}
\hfill
\begin{subfigure}[b]{0.4\textwidth}
\includegraphics[width=\textwidth]{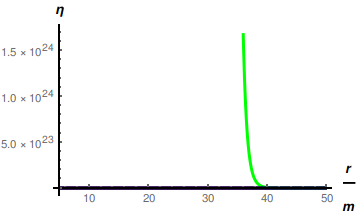} 
\caption{\fcolorbox{red}{red}{\rule{0pt}{2.5pt}\rule{0pt}{2.5pt}} $n_{21}$, \fcolorbox{black}{black}{\rule{0pt}{2.5pt}\rule{0pt}{2.5pt}} $n_{22}$, \fcolorbox{green}{green}{\rule{0pt}{2.5pt}\rule{0pt}{2.5pt}} $n_{23}$, \fcolorbox{blue}{blue}{\rule{0pt}{2.5pt}\rule{0pt}{2.5pt}} $n_{Sch}$} 
\end{subfigure}
\caption{The graph shows the effective refractive indices of some simplified cases of the newly constructed exponential metric Eq.(\ref{(23)}) and (\ref{(52)}) compared to the Schwarzschild metric in the isotropic coordinate. The upper left plot is for the strong field region $r\approx\frac{m}{2}$ and the upper right and lower left plot is for relatively small $r\geq2m$ }
\label{FIG.2}
\end{figure}

Finally, for the metric Eq.(\ref{(78)}), the effective refractive index is 
\begin{equation}
n_3(r)=\exp(\frac{2m}{r}+\alpha^l r^l+\alpha^q r^q +\alpha^n r^n+\alpha^k r^k)
\end{equation}
Again, if $l=q=n=k=0$, then $n_{31}(r)=\exp(\frac{2m}{r}+4)$,
\\if $l=q=0, n=k=-1$, then $n_{32}(r)=\exp(\frac{2m}{r}+\frac{2}{\alpha r}+2)$,
\\if $l=q=0, n=k=1$, then $n_{33}(r)=\exp(\frac{2m}{r}+2\alpha r+2)$,
\\if $l=q=n=k=-1$, then $n_{34}(r)=\exp(\frac{2m}{r}+\frac{4}{\alpha r})$,
\\if $l=q=n=k=1$, then $n_{35}(r)=\exp(\frac{2m}{r}+4\alpha r)$.
\begin{figure}[H]
\begin{subfigure}[b]{0.4\textwidth}
\includegraphics[width=\textwidth]{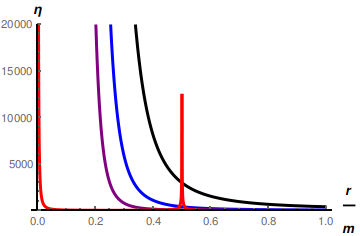}
\caption{\fcolorbox{black}{black}{\rule{0pt}{2.5pt}\rule{0pt}{2.5pt}} $n_{31}$, \fcolorbox{green}{green}{\rule{0pt}{2.5pt}\rule{0pt}{2.5pt}} $n_{32}$, \fcolorbox{blue}{blue}{\rule{0pt}{2.5pt}\rule{0pt}{2.5pt}}$n_{33}$, \fcolorbox{pink}{pink}{\rule{0pt}{2.5pt}\rule{0pt}{2.5pt}} $n_{34}$, \fcolorbox{purple}{purple}{\rule{0pt}{2.5pt}\rule{0pt}{2.5pt}} $n_{35}$, \fcolorbox{red}{red}{\rule{0pt}{2.5pt}\rule{0pt}{2.5pt}} $n_{Sch}$}
\end{subfigure}
\hfill
\begin{subfigure}[b]{0.4\textwidth}
\includegraphics[width=\textwidth]{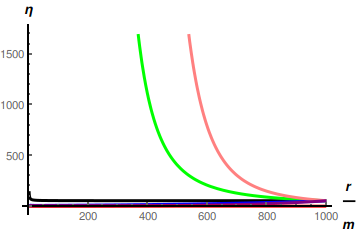} 
\caption{\fcolorbox{black}{black}{\rule{0pt}{2.5pt}\rule{0pt}{2.5pt}} $n_{31}$, \fcolorbox{green}{green}{\rule{0pt}{2.5pt}\rule{0pt}{2.5pt}} $n_{32}$, \fcolorbox{blue}{blue}{\rule{0pt}{2.5pt}\rule{0pt}{2.5pt}}$n_{33}$, \fcolorbox{pink}{pink}{\rule{0pt}{2.5pt}\rule{0pt}{2.5pt}} $n_{34}$, \fcolorbox{purple}{purple}{\rule{0pt}{2.5pt}\rule{0pt}{2.5pt}} $n_{35}$, \fcolorbox{red}{red}{\rule{0pt}{2.5pt}\rule{0pt}{2.5pt}} $n_{Sch}$} 
\end{subfigure}
\caption{The graph shows the effective refractive indices of some simplified cases of the  exponential metric Eq.(\ref{(78)}) compared to the Schwarzschild metric in the isotropic coordinate. The left plot is for the strong field region $r\approx\frac{m}{2}$ and the right plot is for large $r$ }
\label{FIG.4}
\end{figure}
These effective refractive indices specify the focussing/defocussing of null geodesics and they are distinct all the way down to $r=0$. The effective refractive index for the gravitational field in weak and strong field limit was studied in \cite{boonserm2005effective, barcelo2011analogue, visser2013survey}.
\par From Fig.\ref{FIG.1},\ref{FIG.2},\ref{FIG.4} we can see that the effective refractive index for the modified exponential metric Eq.(\ref{(2)}),(\ref{(23)}),(\ref{(52)}),(\ref{(78)}) is slightly greater than that of the Schwarzschild metric in isotropic coordinate at small $r\geq 2m$. As $r$ increases, the curves converge to each other. In the strong field region that is $r\leq \frac{m}{2}$ the curve of the effective refractive index of Schwarzschild metric differ entirely with that of the exponential metrics Eq.(\ref{(2)}),(\ref{(23)}),(\ref{(52)}),(\ref{(78)}). If we stick to the fact that the refractive index is the ratio between the speed of light in vacuum and its phase velocity in a medium, the increase in the refractive index indicates the slowing down of the phase velocity  the phase velocity. Which would signify that light is not able to passthrough the medium. This is exactly what happens for the refractive index  for Schwarzschild metric in Eq.(\ref{sch}), when the isotropic coordinates approaches the event horizon. For an observer at space-like infinity the event horizon acts like a medium with infinite refractive index. even though a light signal can reach the event horizon in a proper time, a distant observer would instead come to the conclusion that the same signal takes an infinite time to cross the event horizon. That is once we get close to the Schwarzschild horizon, the lensing properties are very much different. We can see a clear spike in the curve of Schwarzschild effective refractive index near the event horizon from the above figures. Whereas the effective refractive indices of the exponential metrics are continous in the whole region and they decrease asymptotically as $r\to \infty$.

\section{Photon sphere and innermost stable circular orbit(ISCO)}
In the region of the photon sphere the gravity is enormously strong that even the photons are bound to travel in orbits, thet is why they are also called the last photon orbit. A photon sphere has been defined as a timelike hypersurface of the form $\{r=r_{cl}\}$ where $r_{cl}$ is the closest distance of approach for which the Einstein bending angle of a light ray is large \cite{virbhadra2000schwarzschild}. It is located farther from the centre of a blackhole than the event horizon. There are no stable free fall orbits that exist within or cross the photon sphere. The Schwarzschild unstable circular photon orbit or photon sphere for massless particle is at $r_s=\frac{3GM}{c^2}=\frac{3r_0}{2}=3m$ \cite{claudel2001geometry}, where $r_0$ is the Schwarzschild radius or the radius of the event horizon, $G$ is universal Gravitational constant, $M$ is the mass of the Schwarzschild blackhole and $c$ is the speed of light in vacuum. The existence of a photon sphere in space-time has important implications for gravitational lensing. In any space-time containing a photon sphere, gravitational lensing give rise to relativistic images \cite{virbhadra2000schwarzschild, virbhadra2022distortions, virbhadra2022compactness, adler2022cosmological, virbhadra2009relativistic, virbhadra2008time, virbhadra1998role, virbhadra2002gravitational}. 
\\On the otherhand ISCO(innermost stable circular orbit) is the smallest marginally stable circular orbit in which stable circular motion is still possible. The location of ISCO or the ISCO radius depends on the angular momentum (spin) of the central object \cite{misner1973freeman}. We want to highlight that the notion of ISCO depends only on the geodesic equations, not on the assumed field equations chosen for setting up the spacetime. For a non-spinning massive object, the ISCO radius for Schwarzschild spacetime is $r_s=3r_0=6m$ \cite{jefremov2015innermost, kaplan2022circular, landau2013classical}.
\\Now, in order to find the circular orbit for the modified metric Eq.(\ref{(2)}), let us consider the affinely parameterized tangent vector to the worldline of a massive or massless particle,
\begin{equation}\label{previous}
g_{ab}\frac{dx^a}{d\lambda}\frac{dx^b}{d\lambda}=-e^{-\frac{2m}{r}}\left(\frac{dt}{d\lambda}\right)^2+e^{\frac{2m}{r}+2\alpha r}\left[\left(\frac{dr}{d\lambda}\right)^2+r^2 \left[\left(\frac{d\theta}{d\lambda}\right)^2+sin^2\theta \left(\frac{d\phi}{d\lambda}\right)^2 \right] \right]=\epsilon.
\end{equation}
Where $\epsilon$ can takes value $-1$ or $0$. $-1$ corresponds to a timelike trajectory and $0$ corresponds to a null trajectory. Now we set $\theta=\frac{\pi}{2}$, then the Eq.(\ref{previous}) reduces to,
\begin{equation}\label{above}
g_{ab}\frac{dx^a}{d\lambda}\frac{dx^b}{d\lambda}=-e^{-\frac{2m}{r}}\left(\frac{dt}{d\lambda}\right)^2+e^{\frac{2m}{r}+2\alpha r}\left[\left(\frac{dr}{d\lambda}\right)^2+r^2 \left(\frac{d\phi}{d\lambda}\right)^2 \right]=\epsilon.
\end{equation}
Now, the Killing symmetries imply two conserved quantities, that is energy and angular momentum.
\begin{equation}
e^{-\frac{2m}{r}-\alpha r}\left(\frac{dt}{d\lambda}\right)=E; e^{\frac{2m}{r}+2\alpha r}r^2 \left(\frac{d\phi}{d\lambda}\right)=L.
\end{equation}
Therefore the Eq.(\ref{above}) simplifies as,
\begin{equation}
e^{\frac{2m}{r}+2\alpha r}\left[-E^2+\left(\frac{dr}{d\lambda}\right)^2\right]+e^{-\frac{2m}{r}-2\alpha r}\frac{L^2}{r^2}=\epsilon.
\end{equation}
That is,
\begin{equation}
\left(\frac{dr}{d\lambda}\right)^2=E^2+e^{-\frac{2m}{r}-2\alpha r}\left[\epsilon-e^{-\frac{2m}{r}-2\alpha r}\frac{L^2}{r^2}\right].
\end{equation}
The effective potential for the geodesic orbits is,
\begin{equation}
V_{\epsilon}(r)=e^{-\frac{2m}{r}-2\alpha r}\left[-\epsilon+e^{-\frac{2m}{r}-2\alpha r}\frac{L^2}{r^2} \right].
\end{equation}
\textbf{Case 1} For massless particles such as photons, $\epsilon=0$, then the effective potential for photon is,
\begin{equation}
V_0(r)=\frac{e^{-\frac{4m}{r}-4\alpha r}L^2}{r^2}.
\end{equation}
Now, for finding the maximum value $\frac{dV_0(r)}{dr}=0$, therefore,
\begin{equation}
\frac{dV_0(r)}{dr}=e^{-\frac{4m}{r}-4\alpha r}\left(\frac{4mL^2}{r^4}-\frac{4L^2\alpha}{r^2}-\frac{2L^2}{r^3}\right)=0.
\end{equation}
Which leads to $r=\frac{-1+\sqrt{1+16\alpha m}}{4\alpha}$. Now for $0<m<<\frac{1}{16\alpha}$, we can consider $r=2m$ which corresponds to $V_{0,max}=\frac{L^2}{(2me^{1+4\alpha m})^2}$. Again since $0<m<<\frac{1}{16\alpha}$, we can neglect $4\alpha m$ as it is very very less than $1$. So, the maximum value of $V_0$ is approximately $V_{0,max}=(\frac{L}{2me})^2$. So, we can say that there is a unstable photon sphere at $r=2m$ and in curvature cordinate $r_s=r e^{\frac{1}{2}+2m\alpha}\approx 3.313m$.
\\
\textbf{Case 2} For massive particles such as electrons, protons, atoms or planets, $\epsilon=-1$, the effective potential becomes,
\begin{equation}
V_1(r)=e^{-\frac{2m}{r}-2\alpha r}\left[1+e^{-\frac{2m}{r}-2\alpha r}\frac{L^2}{r^2} \right].
\end{equation}
Now, the first and second derivative of $V_1$ with respect to $r$ are,
\begin{equation}\label{v1}
V_1^/(r)=-\frac{2e^{-\frac{2m}{r}-2\alpha r}}{r^4}[r^2(-m+r^2\alpha)+e^{-\frac{2m}{r}-2\alpha r}L^2(-2m+r+2r^2\alpha)],
\end{equation}
\begin{align}
\begin{split}
V_1^{//}={}&\frac{2e^{-\frac{2m}{r}-2\alpha r}}{r^6}[2r^2(m^2+r^4\alpha^2-mr(1+2\alpha r))+L^2e^{-\frac{2m}{r}-2\alpha r}(8m^2-4mr(3+4\alpha r)\\
&+r^2(3+8\alpha r(1+r\alpha)))].
\end{split}
\end{align}
The circular orbits, $r_c$, occurs at $V_1^/(r_c)=0$. But the determination of $r_c$ as a function of $m$ and $L$ is way more complex. Let us consider the required angular momentum $L_c$ as a function of $r_c$ and $m$. So, from the Eq.(\ref{v1}), we will get,
\begin{equation}
L_c(r_c,m)=\frac{r_ce^{\frac{m}{r_c}+r_c\alpha}\sqrt{m-r_c^2\alpha}}{\sqrt{2r_c^2\alpha +r_c-2m}}.
\end{equation}
Now, the first derivative $L_c(r_c,m)$ with respect to $r_c$ is
\begin{align}
\begin{split}
\frac{\delta L_c(r_c,m)}{\delta r_c}={}&\frac{e^{\frac{m}{r_c}+r_c\alpha}}{2r_c(2r_c^2\alpha +r_c-2m)^{\frac{3}{2}}\sqrt{m-r_c^2\alpha}}[-6m^2r_c+4m^3+mr_c^2-3r_c^4\alpha-12r_c^2m^2\alpha\\
&+12m\alpha r_c^3+12mr_c^4\alpha^2-6r_c^5\alpha^2-4r_c^6\alpha^3].
\end{split}
\end{align}
From this, we can say that $L_c(r_c,m)$ has minimum at $r_c=5.469m$ (since for the metric Eq(2), we have considered the range of $m$ as $0<m<<\frac{1}{16\alpha}$) and the minimum of value of $L_c(r_c,m)$ is $L_{min}=3.4618m$.
\\So, the location of ISCO for massive particle(for the metric Eq.(\ref{(2)}) in isotropic coordinate) is
\begin{equation}
r_{ISCO}\approx5.469m
\end{equation}
and in curvature coordinates, the location is,
\begin{equation}
r_{s,ISCO}=re^{\frac{m}{r}+\alpha r}\approx6.6023m.
\end{equation}
The same result is also valid for the metric Eq.(\ref{(23)}) and Eq.(\ref{(32)}).
\section{Regge-Wheeler equation}
Regge-Wheeler-Zerilli equations are a pair of equations that demonostrates the gravitational perturbations of Schwarzschild black hole in general relativity. The perturbations is divided into two parts, namely, axial and polar perturbations. The equation which includes axial perturbation is called Regge-Wheeler \cite{regge1957stability} equation whereas the equation for polar perturbations is called Zerilli \cite{zerilli1970effective} equation.
\\
Now, we will consider the Regge-Wheeler equation for scalar and vector perturbations around the modified exponential metrics Eq.(\ref{(2)}),(\ref{(78)}) by using inverse Cowling approximation \cite{boonserm2013regge, boonserm2008bounding}.
\\For metric Eq.(\ref{(2)}) 
\begin{equation}
ds^2=-e^{-\frac{2m}{r}}dt^2+e^{\frac{2m}{r}+2\alpha r}[dr^2+r^2d\theta^2+r^2sin^2\theta d\phi^2].
\end{equation}
Let us define a tortoise coordinate $dr_*=e^{\frac{2m}{r}+\alpha r}dr$, then the above equation can be written as,
\begin{equation}
ds^2=e^{-\frac{2m}{r}}[-dt^2+dr_*^2]+e^{\frac{2m}{r}+2\alpha r}r^2[d\theta^2+sin^2\theta d\phi^2].
\end{equation}
$r$ is now an implicit function of $r_*$. So, the above Equation simplifies as,
\begin{equation}
ds^2=e^{-\frac{2m}{r}}[-dt^2+dr_*^2]+r_s^2[d\theta^2+sin^2\theta d\phi^2].
\end{equation}
The Regge-Wheeler equation can be written as \cite{boonserm2013regge, boonserm2008bounding},
\begin{equation}
\delta_*^2\hat{\phi}+(\omega^2-V)\hat{\phi}=0.
\end{equation}
Where $\delta_*$ stands for $\delta_{r_*}$. Now for a spherically symmetric metric represented in curvature coordinates, the Regge-Wheeler potential for spins $S\in{(0,1,2)}$ and angular momentum $l\geq S$ \cite{boonserm2013regge} is,
\begin{equation}
V_s=(-g_{tt})\left[\frac{l(l+1)}{r_s^2}+\frac{S(S-1)(g^{rr}-1)}{r_s^2}\right]+(1-S)\frac{\delta_*^2r_s}{r_s}.
\end{equation}
The metric Eq.(\ref{(2)}) in curvature coordinate is represented by the Eq.(\ref{(104)}). In our case $g_{tt}=-e^{-\frac{2m}{r}}$,  $g^{rr}=(1-\frac{m}{r}+\alpha r)^2$ and $r_s=re^{\frac{m}{r}+\alpha r}$. The value of $\frac{\delta_*^2r_s}{r_s}$ comes out as,
\begin{equation}
\frac{\delta_*^2r_s}{r_s}=\frac{e^{-\frac{4m}{r}-2\alpha r}}{r^4}[2mr-m^2+\alpha r^3+\alpha m r^2].
\end{equation}
Therefore, the Regge-Wheeler potential for the metric Eq.(\ref{(2)}) is,
\begin{align}
\begin{split}
V_s={}&e^{-\frac{4m}{r}-2\alpha r}[\frac{l(l+1)}{r^2}+(S-1)[\frac{m^2(S+1)}{r^4}-\frac{2m(S+1)}{r^3}+\frac{\alpha(2S-1)}{r}\\
&-\frac{\alpha m(2S+1)}{r^2}+S\alpha^2]].
\end{split}
\end{align}
This potential is zero both at $r=0$ and $r=\infty$, with some extreme values at non-trivial values of $r$.
\\
\textbf{1. Spin zero ($S=0$)}
For spin zero, the Regge-Wheeler potential becomes,
\begin{equation}
V_0=e^{-\frac{4m}{r}-2\alpha r}\left[\frac{l(l+1)}{r^2}-\frac{m^2}{r^4}+\frac{2m}{r^3}+\frac{\alpha}{r}+\frac{\alpha m}{r^2}\right].
\end{equation}
Now, for the s-wave($l=0$) the potential simplifies as,
\begin{equation}
V_{0,l=0}=e^{-\frac{4m}{r}-2\alpha r}\left[-\frac{m^2}{r^4}+\frac{2m}{r^3}+\frac{\alpha}{r}+\frac{\alpha m}{r^2}\right],
\end{equation}
\begin{figure}[H]
\begin{subfigure}[b]{0.4\textwidth}
\includegraphics[width=\textwidth]{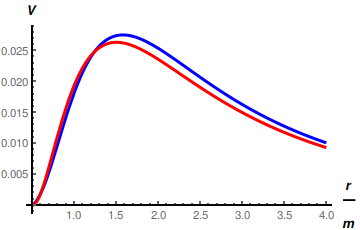}
\caption{\fcolorbox{blue}{blue}{\rule{0pt}{2.5pt}\rule{0pt}{2.5pt}} Exponential metric Eq.(\ref{(2)}), \fcolorbox{red}{red}{\rule{0pt}{2.5pt}\rule{0pt}{2.5pt}} Schwarzschild metric}
\end{subfigure}
\hfill
\begin{subfigure}[b]{0.4\textwidth}
\includegraphics[width=\textwidth]{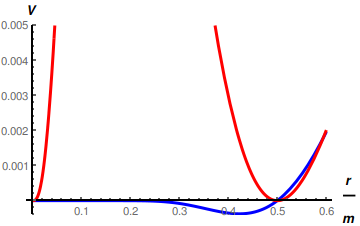} 
\caption{\fcolorbox{blue}{blue}{\rule{0pt}{2.5pt}\rule{0pt}{2.5pt}} Exponential metric Eq.(\ref{(2)}), \fcolorbox{red}{red}{\rule{0pt}{2.5pt}\rule{0pt}{2.5pt}} Schwarzschild metric} 
\end{subfigure}
\caption{The graph depicts the spin zero Regge-Wheeler potential for $l=0$. The potentials are almost similar for $r>\frac{m}{2}$, but they are different for $r<\frac{m}{2}$. For the exponential metric we have considered $0<m<<\frac{1}{4\alpha}$ and $\alpha=\frac{1}{1000}$}
\label{FIG.5}
\end{figure}
Both the potentials have zeroes at $r=\frac{1}{2}m=0.5 m$ and that for $r<\frac{m}{2}$ only the Regge-Wheeler potential is of physical interest. From the above figure we can see that the Regge-Wheeler potential for the exponential metric has a maximum at $r\approx 1.6 m$ and a trough(local minimum) at $r\approx 0.42 m$. On the otherhand the Regge-Wheeler potential for the Schwarzschild metric has maximum of $r=1.5 m$ and another peak at $r\approx 0.166 m$ 
\\
\textbf{2. Spin one ($S=1$)}
For spin one, the Regge-Wheeler potential becomes very simple,($l\geq 1$)
\begin{equation}
V_1=e^{-\frac{4m}{r}-2\alpha r}\frac{l(l+1)}{r^2},
\end{equation}
\begin{figure}[H]
\begin{subfigure}[b]{0.4\textwidth}
\includegraphics[width=\textwidth]{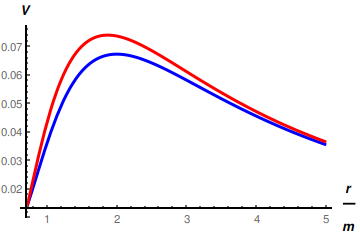}
\caption{\fcolorbox{blue}{blue}{\rule{0pt}{2.5pt}\rule{0pt}{2.5pt}} Exponential metric Eq.(\ref{(2)}), \fcolorbox{red}{red}{\rule{0pt}{2.5pt}\rule{0pt}{2.5pt}} Schwarzschild metric}
\end{subfigure}
\hfill
\begin{subfigure}[b]{0.4\textwidth}
\includegraphics[width=\textwidth]{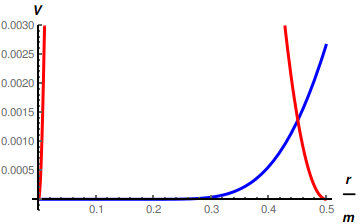} 
\caption{\fcolorbox{blue}{blue}{\rule{0pt}{2.5pt}\rule{0pt}{2.5pt}} Exponential metric Eq.(\ref{(2)}), \fcolorbox{red}{red}{\rule{0pt}{2.5pt}\rule{0pt}{2.5pt}} Schwarzschild metric} 
\end{subfigure}
\caption{The graph depicts the spin one Regge-Wheeler potential for $l=1$. The potentials are almost similar for $r>\frac{m}{2}$, but they are different for $r<\frac{m}{2}$. For the exponential metric we have considered $0<m<<\frac{1}{4\alpha}$ and $\alpha=\frac{1}{1000}$ .}
\label{FIG.6}
\end{figure}
From the above figure we can see that Regge-Wheeler potential for the exponential metric rises from zero to some maximum value at $r\approx 1.92 m$ and then decays back to zero(as $r \to \infty$). Whereas the Potential for the Schwarzschild metric rises from zero to a peak(not maxima) at $r\approx 0.13 m$ then falls to zero at $r=0.5m$ and then again rises to a maximum value at $r\approx 1.86m$ and decays back to zero.
\\
\textbf{3. Spin two ($S=2$)} For spin two axial perturbations, the Regge-Wheeler potential is,($l\geq 2$)
\begin{equation}
V_2=e^{-\frac{4m}{r}-2\alpha r}\left[\frac{l(l+1)}{r^2}+\frac{3m^2}{r^4}-\frac{6m}{r^3}+\frac{3\alpha}{r}-\frac{5\alpha m}{r^2}+2\alpha^2 \right],
\end{equation}
\begin{figure}[H]
\begin{subfigure}[b]{0.4\textwidth}
\includegraphics[width=\textwidth]{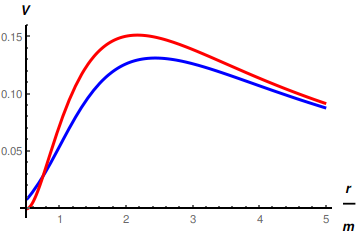}
\caption{\fcolorbox{blue}{blue}{\rule{0pt}{2.5pt}\rule{0pt}{2.5pt}} Exponential metric Eq.(\ref{(2)}), \fcolorbox{red}{red}{\rule{0pt}{2.5pt}\rule{0pt}{2.5pt}} Schwarzschild metric}
\end{subfigure}
\hfill
\begin{subfigure}[b]{0.4\textwidth}
\includegraphics[width=\textwidth]{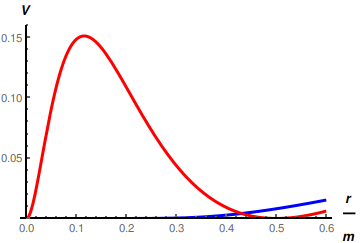} 
\caption{\fcolorbox{blue}{blue}{\rule{0pt}{2.5pt}\rule{0pt}{2.5pt}} Exponential metric Eq.(\ref{(2)}), \fcolorbox{red}{red}{\rule{0pt}{2.5pt}\rule{0pt}{2.5pt}} Schwarzschild metric} 
\end{subfigure}
\caption{The graph depicts the spin two Regge-Wheeler potential for $l=2$. The potentials are almost similar for $r>\frac{m}{2}$, but they are different for $r<\frac{m}{2}$. For the exponential metric we have considered $0<m<<\frac{1}{4\alpha}$ and $\alpha=\frac{1}{1000}$.}
\label{FIG.7}
\end{figure}
Regge-Wheeler potential for the exponential metric rises from zero to some maximum at $r\approx 2.42m$ and then decays back to zero as $r\to\infty$. For the Schwarzschild metric, \cite{boonserm2013regge} Regge-Wheeler potential rises from zero to some maximum at $r\approx 0.18m$ and falls back to zero at $r=0.5m$ then again it rises and attains a peak value at $r\approx 2.186 m$(not a maxima) and then decays back to zero as $r\to\infty$.
\\Now, consider the Regge-Wheeler potential for Schwarzschild spacetime,
\begin{equation}
V_{s,Sch}=\left(\frac{1-\frac{m}{2r}}{1+\frac{m}{2r}}\right)^2\left[\frac{l(l+1)}{r^2(1+\frac{m}{2r})^4}-\frac{(S^2-1)2m}{r^3(1+\frac{m}{2r})^6} \right].
\end{equation}
For, spin zero s-wave, the potential simplifies as,
\begin{equation}
V_{0,l=0,Sch}=\left(\frac{1-\frac{m}{2r}}{1+\frac{m}{2r}}\right)^2\left[\frac{2m}{r^3(1+\frac{m}{2r})^6} \right].
\end{equation}
Similarly for spin one and spin two \cite{boonserm2013regge}, the potential turns out as,
\begin{equation}
V_{1,Sch}=\frac{(1-\frac{m}{2r})^2}{(1+\frac{m}{2r})^6}\frac{l(l+1)}{r^2}
\end{equation}
and
\begin{equation}
V_{2,Sch}=\left(\frac{1-\frac{m}{2r}}{1+\frac{m}{2r}} \right)^2\left[\frac{l(l+1)}{r^2(1+\frac{m}{2r})^4}-\frac{6m}{r^3(1+\frac{m}{2r})^6} \right].
\end{equation}
respectively.
\\
Now, consider the metric represented by Eq.(\ref{(78)})
\begin{equation}
ds^2 = -\exp\left(-\frac{2m}{r}-2\alpha^l r^l-2\alpha^q r^q\right)dt^2 + \exp\left(\frac{2m}{r}+2\alpha^n r^n+2\alpha^k r^k\right)[dr^2 + r^2 d\theta^2 + r^2 sin^2\theta d\phi^2],
\end{equation}
in curvature corrdinates, the above metric takes the form,
\begin{equation}
ds^2=-\exp\left(-\frac{2m}{r}-2\alpha^l r^l-2\alpha^q r^q\right)dt^2+\frac{1}{(1-\frac{m}{r}+n\alpha^n r^n+k\alpha^k r^k)^2}dr_s^2+ r_s^2(d\theta^2+sin^2\theta d\phi^2).
\end{equation}
The tortoise coordinate for this metric is defined as,
\begin{equation}
dr_*=e^{\frac{2m}{r}+\alpha^nr^n+\alpha^kr^k+\alpha^lr^l+\alpha^qr^q}.
\end{equation}
Then the metric takes the form,
\begin{equation}
ds^2=e^{-\frac{2m}{r}-2\alpha^lr^l-2\alpha^qr^q}(-dt^2+dr_*^2)+r_s^2(d\theta^2+sin^2\theta d\phi^2).
\end{equation}
Therefore, the Regge-Wheeler potential for this metric comes out as,
\begin{align}
\begin{split}
V_s={}&e^{(-\frac{4m}{r}-2\alpha^nr^n-2\alpha^kr^k-2\alpha^lr^l-2\alpha^qr^q)}[\frac{l(l+1)}{r^2}+\frac{S(S-1)}{r^2}[(1-\frac{m}{r}+n\alpha^nr^n+k\alpha^kr^k)^2\\
&-1]+(1-S)[\frac{2m}{r^3}-\frac{m^2}{r^4}+n^2\alpha^nr^{n-2}+k^2\alpha^kr^{k-2}-l\alpha^lr^{l-2}-q\alpha^qr^{q-2}+ml\alpha^lr^{l-3}\\
&+mq\alpha^qr^{q-3}+mn\alpha^nr^{n-3}-ln\alpha^{n+l}r^{n+l-2}-qn\alpha^{n+q}r^{n+q-2}+mk\alpha^kr^{k-3}-lk\alpha^{k+l}r^{k+l-2}-\\
&qk\alpha^{k+q}r^{k+q-2}]].
\end{split}
\end{align}
\textbf{1. Spin zero(S=0)} For zero spin $S=0$, for s-wave $l=0$, then the Regge-Wheeler potnetial simplifies as,
\begin{align}
\begin{split}
V_0={}&e^{(-\frac{4m}{r}-2\alpha^nr^n-2\alpha^kr^k-2\alpha^lr^l-2\alpha^qr^q)}[\frac{2m}{r^3}-\frac{m^2}{r^4}+n^2\alpha^nr^{n-2}+k^2\alpha^kr^{k-2}-l\alpha^lr^{l-2}-q\alpha^qr^{q-2}\\
&+ml\alpha^lr^{l-3}+mq\alpha^qr^{q-3}+mn\alpha^nr^{n-3}-ln\alpha^{n+l}r^{n+l-2}-qn\alpha^{n+q}r^{n+q-2}+mk\alpha^kr^{k-3}\\
&-lk\alpha^{k+l}r^{k+l-2}-qk\alpha^{k+q}r^{k+q-2}].
\end{split}
\end{align}
Now, let us consider some simple cases of this potential for various values of $n,k,l$ and $q$.
\\
\textbf{if $n=k=l=q=0$} then the potential takes the form
\begin{equation}
V_{01}=e^{-\frac{4m}{r}-8}[\frac{2m}{r^3}-\frac{m^2}{r^4}].
\end{equation}
\textbf{if $n=k=-1$, $l=q=0$ or $n=k=0$, $l=q=-1$} then we will get,
\begin{equation}
V_{02}=e^{-\frac{4m}{r}-\frac{4}{\alpha r}-4}[\frac{2m}{r^3}-\frac{m^2}{r^4}+\frac{2}{\alpha r^3}-\frac{2 m}{\alpha r^4}].
\end{equation}
\textbf{if $n=k=1$ and $l=q=0$} then
\begin{equation}
V_{03}=e^{-\frac{4m}{r}-4\alpha r-4}[\frac{2m}{r^3}-\frac{m^2}{r^4}+\frac{2\alpha}{r}+\frac{2m\alpha}{r^2}].
\end{equation}
\textbf{if $l=n=0$ and $q=k=-1$ or $q=k=0$ and $l=n=-1$} then the potential comes out as,
\begin{equation}
V_{04}=e^{-\frac{4m}{r}-\frac{4}{\alpha r}-4}[\frac{2m}{r^3}-\frac{m^2}{r^4}+\frac{2}{\alpha r^3}-\frac{2 m}{\alpha r^4}-\frac{1}{\alpha^2 r^4}].
\end{equation}
\textbf{if $l=n=0$ and $q=k=1$ or $q=k=0$ and $l=n=1$} then
\begin{equation}
V_{05}=e^{-\frac{4m}{r}-4\alpha r-4}[\frac{2m}{r^3}-\frac{m^2}{r^4}+\frac{2\alpha m}{r^2}-\alpha^2].
\end{equation}
\textbf{if $l=q=1$ and $n=k=0$} the potential becomes,
\begin{equation}
V_{06}=e^{-\frac{4m}{r}-4r\alpha-4}[\frac{2m}{r^3}-\frac{m^2}{r^4}+\frac{2m\alpha}{r^2}-\frac{2\alpha}{r}].
\end{equation}
\textbf{if $l=q=n=k=1$} the Regge-Wheeler potential takes the form,
\begin{equation}
V_{07}=e^{-\frac{4m}{r}-8\alpha r}[\frac{2m}{r^3}-\frac{m^2}{r^4}+\frac{4m\alpha}{r^2}-4\alpha^2].
\end{equation}
\begin{figure}[H]
\begin{subfigure}[b]{0.4\textwidth}
\includegraphics[width=\textwidth]{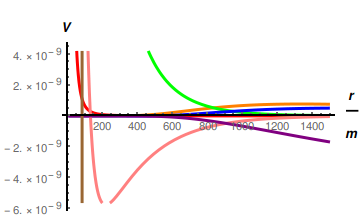}
\caption{\fcolorbox{red}{red}{\rule{0pt}{2.5pt}\rule{0pt}{2.5pt}} $V_{01}$, \fcolorbox{orange}{orange}{\rule{0pt}{2.5pt}\rule{0pt}{2.5pt}} $V_{02}$, \fcolorbox{green}{green}{\rule{0pt}{2.5pt}\rule{0pt}{2.5pt}} $V_{03}$, \fcolorbox{blue}{blue}{\rule{0pt}{2.5pt}\rule{0pt}{2.5pt}} $V_{04}$, \fcolorbox{pink}{pink}{\rule{0pt}{2.5pt}\rule{0pt}{2.5pt}} $V_{05}$, \fcolorbox{purple}{purple}{\rule{0pt}{2.5pt}\rule{0pt}{2.5pt}} $V_{06}$, \fcolorbox{brown}{brown}{\rule{0pt}{2.5pt}\rule{0pt}{2.5pt}} $V_{07}$}
\end{subfigure}
\hfill
\begin{subfigure}[b]{0.4\textwidth}
\includegraphics[width=\textwidth]{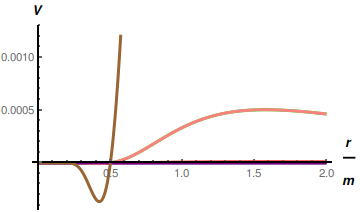} 
\caption{\fcolorbox{red}{red}{\rule{0pt}{2.5pt}\rule{0pt}{2.5pt}} $V_{01}$, \fcolorbox{orange}{orange}{\rule{0pt}{2.5pt}\rule{0pt}{2.5pt}} $V_{02}$, \fcolorbox{green}{green}{\rule{0pt}{2.5pt}\rule{0pt}{2.5pt}} $V_{03}$, \fcolorbox{blue}{blue}{\rule{0pt}{2.5pt}\rule{0pt}{2.5pt}} $V_{04}$, \fcolorbox{pink}{pink}{\rule{0pt}{2.5pt}\rule{0pt}{2.5pt}} $V_{05}$, \fcolorbox{purple}{purple}{\rule{0pt}{2.5pt}\rule{0pt}{2.5pt}} $V_{06}$, \fcolorbox{brown}{brown}{\rule{0pt}{2.5pt}\rule{0pt}{2.5pt}} $V_{07}$} 
\end{subfigure}
\caption{The graph depicts the spin zero Regge-Wheeler potential for $l=0$ various cases. The shapes of the graphs are almost similar for the different ranges of $r$. }
\label{FIG.8}
\end{figure}
\textbf{2. Spin one($S=1$)} For spin one, the Regge-Wheeler takes the form,
\begin{equation}
V_1=e^{-\frac{4m}{r}-2\alpha^nr^n-2\alpha^kr^k-2\alpha^lr^l-2\alpha^qr^q}[\frac{l(l+1)}{r^2}]
\end{equation}
In this potential, $n,k,l$ and $q$ have equal footing. Depending on their values, let us consider some simplest cases.
\\
If all of them are zero, then $V_{11}=e^{-\frac{4m}{r}-8}[\frac{l(l+1)}{r^2}]$,
\\
If two of them is negative one and the remaining two's are zero, then $V_{12}=e^{-\frac{4m}{r}-\frac{4}{\alpha r}-4}[\frac{l(l+1)}{r^2}]$,
\\
If two of them are one and the remaining two's are zero, then $V_{13}=e^{-\frac{4m}{r}-4\alpha r-4}[\frac{l(l+1)}{r^2}]$,
\\
If all of them are negative one, then $V_{14}=e^{-\frac{4m}{r}-\frac{8}{\alpha r}}[\frac{l(l+1)}{r^2}]$,
\\
If all of them are one, then $V_{15}=e^{-\frac{4m}{r}-8\alpha r}[\frac{l(l+1)}{r^2}]$.
\begin{figure}[H]
\begin{subfigure}[b]{0.4\textwidth}
\includegraphics[width=\textwidth]{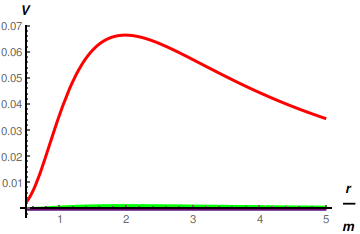}
\caption{\fcolorbox{blue}{blue}{\rule{0pt}{2.5pt}\rule{0pt}{2.5pt}} $V_{11}$, \fcolorbox{black}{black}{\rule{0pt}{2.5pt}\rule{0pt}{2.5pt}} $V_{12}$, \fcolorbox{green}{green}{\rule{0pt}{2.5pt}\rule{0pt}{2.5pt}} $V_{13}$, \fcolorbox{purple}{purple}{\rule{0pt}{2.5pt}\rule{0pt}{2.5pt}} $V_{14}$, \fcolorbox{red}{red}{\rule{0pt}{2.5pt}\rule{0pt}{2.5pt}} $V_{15}$}
\end{subfigure}
\hfill
\begin{subfigure}[b]{0.4\textwidth}
\includegraphics[width=\textwidth]{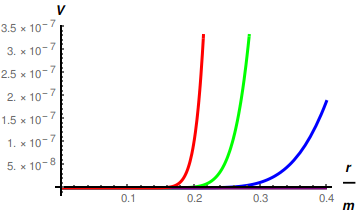} 
\caption{\fcolorbox{blue}{blue}{\rule{0pt}{2.5pt}\rule{0pt}{2.5pt}} $V_{11}$, \fcolorbox{black}{black}{\rule{0pt}{2.5pt}\rule{0pt}{2.5pt}} $V_{12}$, \fcolorbox{green}{green}{\rule{0pt}{2.5pt}\rule{0pt}{2.5pt}} $V_{13}$, \fcolorbox{purple}{purple}{\rule{0pt}{2.5pt}\rule{0pt}{2.5pt}} $V_{14}$, \fcolorbox{red}{red}{\rule{0pt}{2.5pt}\rule{0pt}{2.5pt}} $V_{15}$} 
\end{subfigure}
\caption{The graph depicts the spin one Regge-Wheeler potential for $l=1$ various cases. The shapes of the graphs are almost similar for the different ranges of $r$. }
\label{FIG.9}
\end{figure}
From the graphs presented in Fig.\ref{FIG.5},\ref{FIG.6},\ref{FIG.7},\ref{FIG.8},\ref{FIG.9} we can clearly see that the Regge-Wheeler potentials of the modified exonential metrics Eq.(\ref{(2)}),(\ref{(78)}) for spin zero($s=0$), spin one($s=1$) and spin two($s=2$) are almost similar to that of the Schwarzschild metric for $r\geq \frac{m}{2}$, but they are notably different for $r\leq \frac{m}{2}$. The Regge-Wheeler potential for Schwarzschild is only formal since one is behind a horizon(Schwarzschild metric has a horizon at $r=\frac{m}{2}$) and cannot interact with the domain of outer communication\cite{boonserm2018exponential}.

\section{Results and discussion}
\par In this paper we have focused in forming a series of traversable exponential wormhole metric in isotropic coordinates. The metrics we have formulated here can be viewed as a theoretical model. Where the  metrics have some interesting properties. They violate the Null Energy Condition(NEC). There are some geometric considerations showing that the wormhole throat, a compact two surface of minimal area can exist if only the NEC is violated.
\par All the formulated metrics are Lorentzian and have no horizons as $g_{tt}\neq0$, with the radius of the throat as $r=m$, $r = \frac{-1+\sqrt{1+4\alpha m}}{2\alpha}$, $r=\frac{-1+\sqrt{1+8\alpha m}}{4\alpha}$, $r=m+\frac{1}{\alpha}$, $r=m+\frac{2}{\alpha}$...etc for different types of special cases. But here we have imposed  constrain conditions that $0<m<<\frac{1}{4\alpha}$ (for the radius $r = \frac{-1+\sqrt{1+4\alpha m}}{2\alpha}$), $0<m<<\frac{1}{8\alpha}$ (for the radius $r=\frac{-1+\sqrt{1+8\alpha m}}{4\alpha}$), $\alpha>>m$(for $r=m+\frac{1}{\alpha}$ and $r=m+\frac{2}{\alpha}$), so that the radius at the throat can be considered as $r=r_0=m$, which are definitely some special cases for different range of $m$. Interestingly we find that all the metric components are finite and diagonal components are non zero at $r=r_0=m$. Again on studying the curvature components and scalar invariants of all the formulated metrics,  we observed that they are finite everywhere in the exponential spacetime, most importantly they are finite at the throat and goes to zero both as $r\to\infty$ and as $r\to0$, maximum values are obtained only near the throat. These metrics also donot possess any kind of parallelly propagated naked curvature singularity. Therefore they form a new class of traversable wormhole. We have constructed the metrics Eq.(\ref{(14)}), Eq.(\ref{(23)}), Eq.(\ref{(31)}) and Eq.(\ref{(32)}) as the general temporal exponential wormhole metrics(where the spatial component is $\exp(\frac{2m}{r}+2\alpha r)$), the metrics Eq.(\ref{(37)}), Eq.(\ref{(52)}), Eq.(\ref{(69)}) as the general spatial exponential wormhole metric and Eq.(\ref{(70)}), Eq.(\ref{(78)}), Eq.(\ref{(100)}), Eq.(\ref{(101)}) and Eq.(\ref{(102)}) as the general exponential wormhole metric. The radius of the throat remains the same if we keep the spatial component of the metric fixed. But any change in the temporal component will lead to a different throat radius. From the flare-out condition, it is observed that, the value of $\alpha$ should be very small compared to $m$ when the power terms $n$, $l$, $k$ etc are taken ase positive integer, on the otherhand the value of $\alpha$ should be very large when $n$, $l$, $k$ etc are considered as negative integers.

\par The effective refractive indices of these newly constructed metrics are continous in the whole region of spacetime  and they decrease asymptotically as $r\to \infty$. The lensing properties of these metrics are notably different from the Schwarzschild metric in the strong field region. The innermost stable circular orbits and the unstable photon spheres still exist for these modified metrics but locations of ISCO and Photon spheres are slightly shifted from those of Schwarzschild spacetime. Again the Regge-Wheeler potentials for the various values of spin are also found out to be continous, whereas the Regge-Wheeler potential for Schwarzschild is only formal since one is behind a horizon(Schwarzschild metric has a horizon at $r=\frac{m}{2}$) and cannot interact with the domain of outer communication.

\par So from the above discussed factors we can conclude that the metrics we have formulated are some forms of a new class of general traversable exponential wormhole metrics. All the three types of metrics and other metrics which can be obtained from these three types would share same throat for the same spatial component (temporal component may be different) and for same range of values of $m$.
\par All these novel traversable wormhole metrics violate the necessary Null Energy Condition near the throat or we can say that the throats are filled with some exotic matter. It was thought that exotic matters required to maintain their mouths opened. One could saythat the phantom energy could provide an infinite source of exotic matter necessary to comfort wormhole geometries. We can argue that these exponential metrics has a natural explanation in terms of general relativity coupled to a phantom field.
\par Although the calculations of these wormhole metrics are complicated at present, but doing further research, the calculations can be simplified in the near future which will definitely give some important results and make a significant contribution in the theory of wormholes.
\section*{Data availability statement}
There is no data associated with this article.
\section*{Acknowledgement}
This work is supported by University Grants Commission, Ministry of Education, Govt. of India(NFOBC No.F. 82-44/2020(SA-III)) under the scheme NFOBC programme.
\\
\bibliographystyle{ieeetr}
\bibliography{myref}

\end{document}